\documentclass{ws-ijmpb}

\usepackage{color}
\usepackage{hyperref}

\DeclareMathOperator{\sym}{symm}

\DeclareMathOperator{\grad}{grad}

\DeclareMathOperator{\sech}{sech}
\newcommand{\tens}[1]{\boldsymbol{\mathsf{#1}}}

\allowdisplaybreaks

\begin{document}

\markboth{I.C.\ Christov, T.\ Kress \& A. Saxena}
{Peakompactons: Peaked compact nonlinear waves}

%
\catchline{}{}{}{}{}
%

\title{Peakompactons: Peaked compact nonlinear waves}

\author{Ivan C.\ Christov}

\address{School of Mechanical Engineering, Purdue University\\
West Lafayette, IN 47907, USA\\
christov@purdue.edu}

\author{Tyler Kress}

\address{Department of Mathematics, University of North Carolina, Chapel Hill\\
Chapel Hill, NC 27599, USA}

\author{Avadh Saxena}

\address{Theoretical Division and Center for Nonlinear Studies, Los Alamos National Laboratory\\
Los Alamos, NM 87545, USA}

\maketitle

\begin{history}
\received{Day Month Year}
\revised{Day Month Year}
\end{history}

\begin{abstract}
This article is meant as an accessible introduction to/tutorial on the analytical construction and numerical simulation of a class of non-standard solitary waves termed \emph{peakompactons}. These peaked compactly supported waves arise as solutions to nonlinear evolution equations from a hierarchy of nonlinearly dispersive Korteweg--de Vries-type models. Peakompactons, like the now-well-know compactons and unlike the soliton solutions of the Korteweg--de Vries equation, have finite support, {\it i.e.}, they are of finite wavelength. However, unlike compactons, peakompactons are also peaked, {\it i.e.}, a higher spatial derivative suffers a jump discontinuity at the wave's crest. Here, we construct such solutions exactly by reducing the governing partial differential equation to a nonlinear ordinary differential equation and employing a phase-plane analysis. A simple, but reliable, finite-difference scheme is also  designed and tested for the simulation of collisions of peakompactons. In addition to the peakompacton class of solutions, the general physical features of the so-called $K^\#(n,m)$ hierarchy of nonlinearly dispersive Korteweg--de Vries-type models are discussed as well. 
%
\end{abstract}


\section{Introduction}

\emph{Peakompactons}, the definition of which will be made more precise below, are peaked compactly supported solutions to nonlinear evolution equations from a hierarchy of nonlinearly dispersive Korteweg--de Vries-type models. First, we motivate this class of models by reviewing some modern theories of wave propagation in generalized continua.

\subsection{Motivating example: Waves in generalized continua}
\label{sec:motive}

The governing equations for peakompactons originate from the study of waves in certain continua with an inherent length scale. Such \emph{generalized continua}\cite{GenContA,GenContB} arise in the extension of classical continuum mechanics to the modeling of ``real'' materials. For example, myriads of materials ranging from metals to composites to granular media feature a ``minimal'' (often called \emph{inherent}) length scale, be it the atomic lattice spacing, the typical size of inhomogeneities or the average grain diameter. 

In 1995, Rubin {\it et al.}\cite{Rubin1995} posed the question: can a simple model capture dispersion of waves in non-ideal (generalized) continua due to an inherent length scale? Their proposed model, which is nowadays termed ``Rubin--Rosenau--Gottlieb (RRG) theory,'' inherits and satisfies all the thermodynamic restrictions on the admissibility of motions and parameters of the classical thermoviscous compressible Newtonian fluid. As a result, a self-contained constitutive equation can be obtained. Letting the total stress in the material be $\tens{T} = \tens{T}^{(1)} + \tens{T}^{(2)}$, where 
\begin{equation}\label{eq:T1}
	\tens{T}^{(1)} = -\wp\tens{I} + 2\mu\sym[\nabla\bm{u}] + \big(\mu_\mathrm{B} - \tfrac{2}{3}\mu\big)(\nabla\cdot\bm{u})\tens{I}
\end{equation}
is the Cauchy stress of a thermoviscous compressible Newtonian fluid (see, {\it e.g.}, \S15 in Ref.~\refcite{LandauLifshitz}), the dispersive regularization takes the form
\begin{equation}\label{eq:T2}
	\tens{T}^{(2)} = \varrho\Psi\left\{2\sym[\nabla\dot{\bm{u}}] + 2 (\nabla\bm{u})^\top\nabla\bm{u} - 4(\sym[\nabla\bm{u}])^2\right\}.
\end{equation}
In Eqs.~\eqref{eq:T1} and \eqref{eq:T2} above, $\wp$ represents the thermodynamic pressure, $\varrho$ is the fluid's density, $\mu$ and $\mu_\mathrm{B}$ are the fluid's shear and bulk viscosities respectively, $\tens{I}$ is the identity tensor, $\bm{u}$ is the velocity vector, $\sym[\tens{F}] = \tfrac{1}{2}\big(\tens{F} + \tens{F}^\top\big)$ returns the symmetric part of the tensor field $\tens{F}$, a superscript $\top$ denotes transpose, a superimposed dot denotes the material derivative $\partial(\cdot)/\partial t + (\bm{u}\cdot\nabla)(\cdot)$, and $\Psi$ is a parameter introduced in the RRG theory. Rubin {\it et al.}\cite{Rubin1995} term $\Psi$, which is positive and has the units of length squared, the \emph{dispersion function}. It is further assumed that $\Psi$ can depend only on the rate of deformation tensor's second invariant, namely the scalar $\delta = \sym[\grad\bm{u}]:\sym[\grad\bm{u}]$.
	
Constitutive relations of the form $\tens{T} = \tens{T}^{(1)} + \tens{T}^{(2)}$, with $\tens{T}^{(1)}$ and $\tens{T}^{(2)}$ given by Eqs.~\eqref{eq:T1} and \eqref{eq:T2}, share some general features with models such as the so-called second-grade fluid of Coleman \& Noll,\cite{CN60,DF74} the Navier--Stokes-$\alpha$ models in turbulence,\cite{Chen,FHT01,GOP03} Lagrangian-averaged Euler-$\alpha$ models,\cite{Holm1998b,Holm2002,BF06,MKJC11} finite-scale theories,\cite{Margolin2009,Margolin2014,Jordan2015,Margolin2016} etc. Under all of these formulations, the governing system of flow equations is modified to include an inherent length scale, which is meant to model some desired physical effect or, often, attempts to capture (in a simple way) a number of ``subgrid'' (or, unmodeled) effects.

Recently,\cite{Destrade2006,Jordan2012,Rubin2013} it has been of interest to combine the RRG constitutive relation with the usual continuum balances of mass and linear momentum (in the absence of body forces), namely,\cite{LandauLifshitz}
\begin{subequations}
\begin{align}
\dot{\varrho} + \varrho\nabla\cdot\bm{u} &= 0,\\
\varrho\dot{\bm{u}} &= \nabla\cdot\tens{T},
\end{align}
\end{subequations}
and to derive (unidirectional) \emph{nonlinear evolution equations} for waves in these generalized continua. Specifically, by standard methods,\cite{Jeffrey1972,Crighton1979} Destrade \& Saccomandi\cite{Destrade2006} obtained a one-dimensional (1D) dimensionless equation for unidirectional propagation of weakly-nonlinear shear waves in a hyperelastic, incompressible solid, under the assumption that $\Psi \propto (\sym[\nabla\bm{u}])^3$,
\begin{equation}\label{eq:DS}
	v_t + \tfrac{1}{2}(v^3)_x + \tfrac{1}{3}[(v_x)^3]_{xx} = 0,
\end{equation}
where $v$ is a strain. Henceforth, $x$ and $t$ subscripts stand for partial differentiation with respect to $x$ and $t$. (Although solitons are often associated with waves in fluids, and above we only reviewed the RRG theory in the context of fluids, an excellent overview of the history of and context for solitary wave propagation in elastic solids is given by Maugin.\cite{Maugin2011}) Meanwhile, Jordan \& Saccomandi,\cite{Jordan2012} also working in 1D, obtained a dimensionless equation for unidirectional propagation of weakly-nonlinear acoustic waves in an inviscid, non-thermally conducting compressible fluid, under the assumption that $\Psi \propto (\sym[\nabla\bm{u}])^2$,
\begin{equation}\label{eq:JS}
	u_t + \epsilon b u u_x + \tfrac{1}{6}a_1 [(u_x)^3]_{xx} = 0,
\end{equation}
where $u$ is a velocity, $\epsilon$ is a Mach number, $b$ is the so-called coefficient of nonlinearity of the fluid,\cite{Laut2007} and $a_1$ is related to the material dispersion length scale $\sqrt{\Psi}$. Here, it should also be noted, however, that some anomalous behaviors of wave propagation under the RRG theory have been observed when viscosity is taken into account\cite{Jordan2014} [unlike in Eq.~\eqref{eq:JS}, which is derived for \emph{inviscid} fluids]. 
Notice that the relationship between the advective nonlinearities in Eqs.~\eqref{eq:JS} and \eqref{eq:DS}, namely $uu_x$ versus $v^2v_x$, is precisely the same as for the Korteweg--de Vries (KdV) and the \emph{modified} KdV\cite{Miura1968} equations.

Surprisingly, it has been shown recently\cite{C15} that Eqs.~\eqref{eq:DS} and \eqref{eq:JS} belong to a \emph{single hierarchy} of generalized KdV-like nonlinearly dispersive partial differential equations (PDEs) with \emph{Hamiltonian structure}. (It goes without saying that Hamiltonian structure is highly desirable as it underpins all of classical and modern mechanics of both point particles and continua, including wave phenomena and evolution equations.\cite{Holm2009,Morrison1998,Zakharov1997,Olver1980}) In the present context, the KdV equation, first introduced by Boussinesq\cite{Boussinesq1877} and later re-derived and examined in detail by Korteweg \& de Vries,\cite{Korteweg1895} takes the form
\begin{equation}\label{eq:KdV}
u_t + uu_x + u_{xxx} = 0,
\end{equation}
when properly normalized. Since the seminal discovery of elastic interactions of solitons by Zabusky \& Kruskal\cite{Zabusky1965} (see also Refs.~\refcite{Miura1976,Drazin1989}), the KdV equation has been a staple of \emph{nonlinear science}.\cite{Scott2003} The key difference between Eq.~\eqref{eq:KdV} and Eqs.~\eqref{eq:DS} and \eqref{eq:JS} is in the last term on their respective left-hand sides. While, Eq.~\eqref{eq:KdV} features the \emph{linear} dispersive term $u_{xxx}$, Eqs.~\eqref{eq:DS} and \eqref{eq:JS} feature the \emph{nonlinear} dispersive terms $[(v_x)^3]_{xx}$ and $[(u_x)^3]_{xx}$, respectively. As early as 1974, Kruskal\cite{Kruskal1975} pondered the possibility of nonlinearly dispersive evolution equations and their role in soliton theory. Thus, given the importance of the KdV equation in countless fields of mathematics and science, it behooves us to understand such ``straightforward'' (and physically relevant) nonlinearly dispersive extensions of KdV as the ones given in Eqs.~\eqref{eq:DS} and \eqref{eq:JS} above.

\subsection{$K^\#(n,m)$: A nonlinearly dispersive KdV hierarchy}
\label{sec:Ksharp}
To summarize the derivation from Ref.~\refcite{C15}, consider the properly normalized Lagrangian density for the generic $(1+1)$-dimensional field $\varphi = \varphi(x,t)$: 
\begin{equation}\label{eq:Lagr}
	\mathcal{L}(\varphi;x,t) = \frac{1}{2}\varphi_x \varphi_t + \frac{1}{(n+2)(n+1)} (\varphi_x)^{n+2} - \frac{1}{m+1}(\varphi_{xx})^{m+1},\quad n,m>0,
\end{equation}
where the Lagrangian density corresponding to the classical KdV equation is the special case $n=m=1$.\cite{Gardner1971}
The corresponding action functional is $\int \mathrm{d}x\mathrm{d}t\, \mathcal{L}$. It is extremized by requiring that the field $\varphi$ satisfies the Euler--Lagrange equation (see, {\it e.g.}, \S11 and \S35 in Ref.~\refcite{Gelfand2000}):
\begin{equation}\label{eq:EL}
	\frac{\partial \mathcal{L}}{\partial \varphi} - \frac{\partial}{\partial t}\left(\frac{\partial \mathcal{L}}{\partial \varphi_t}\right) - \frac{\partial}{\partial x}\left(\frac{\partial \mathcal{L}}{\partial \varphi_x}\right) + \frac{\partial^2}{\partial x^2}\left(\frac{\partial \mathcal{L}}{\partial \varphi_{xx}}\right) = 0.
\end{equation}
Substituting Eq.~\eqref{eq:Lagr} into Eq.~\eqref{eq:EL}, we obtain the governing nonlinear evolution equation
\begin{equation}\label{eq:Ksharp}
	K^\#(n,m):\quad u_t + u^{n}u_x + \left[(u_x)^{m}\right]_{xx} = 0,\qquad u(x,t) \equiv \varphi_x(x,t),
\end{equation}
where we have introduced the notation ``$K^\#(n,m)$'' to represent this two-parameter family of PDEs.
It is evident that Eq.~\eqref{eq:DS} [$K^\#(2,3)$] and Eq.~\eqref{eq:JS} [$K^\#(1,3)$], which were derived to describe waves in the generalized RRG continua with an inherent length scale, are special cases of Eq.~\eqref{eq:Ksharp}, subject to proper normalization of $u$.\cite{C15}

As expected from Noether's theorem,\cite{Kruskal1966} three conserved quantities ({\it i.e.}, quantities that are \emph{constant} during the \emph{time evolution} of the field $\varphi$) exist\cite{C15} for Hamiltonian equations of the form given in Eq.~\eqref{eq:Ksharp}, namely
\begin{subequations}
\begin{align}
	H &\equiv \int_{-\infty}^{+\infty}\mathrm{d}x\, \mathcal{H} = \int_{-\infty}^{+\infty}\mathrm{d}x\, \left[ -\frac{1}{(n+2)(n+1)} u^{n+2} + \frac{1}{m+1}(u_x)^{m+1}\right],\\
	M &\equiv \int_{-\infty}^{+\infty}\mathrm{d}x\, \frac{\partial \mathcal{L}}{\partial \varphi_t} = \frac{1}{2}\int_{-\infty}^{+\infty}\mathrm{d}x\, \varphi_x = \frac{1}{2}\int_{-\infty}^{+\infty}\mathrm{d}x\, u,\label{eq:M_invar}\\
	P &\equiv -\int_{-\infty}^{+\infty} \mathrm{d}x\, \frac{\partial \mathcal{L}}{\partial \varphi_t}\varphi_x = - \frac{1}{2} \int_{-\infty}^{+\infty} \mathrm{d}x\, (\varphi_x)^2 = -\frac{1}{2} \int_{-\infty}^{+\infty} \mathrm{d}x\, u^2.\label{eq:P_invar}
\end{align}\label{eq:HMP}%
\end{subequations}
Here, $H$, $M$ and $P$ denote the Hamiltonian (total energy), total wave mass and total wave momentum\cite{Maugin2002} associated with the field $\varphi$. Note that the Hamiltonian density $\mathcal{H}$ is found via the Legendre transformation,\cite{LandauLifshitz2} namely $\mathcal{H} =\frac{\partial \mathcal{L}}{\partial \varphi_t}\varphi_t - \mathcal{L}$. The three quantities in Eq.~\eqref{eq:HMP} are conserved as a result of the translational invariance in space ($x \mapsto x + x_0$ for any constant $x_0$), translation invariance in time ($t \mapsto t + t_0$ for any constant $t_0$), and shift invariance of the field ($\varphi \mapsto \varphi + \varphi_0$ for any constant $\varphi_0$). It has been shown\cite{C15} that \emph{no} other conserved quantities of the form $\int_{-\infty}^{+\infty} \mathrm{d}x\, u^k$, beyond $M$ and $P$ given in Eqs.~\eqref{eq:M_invar} and \eqref{eq:P_invar}, exist for Eq.~\eqref{eq:Ksharp}.

The remaining details of the canonical structure corresponding to such Hamiltonian PDEs (including Poisson brackets, etc.) can be built-up by following the derivations by Cooper {\it et al.}\cite{Cooper1993} (see also the discussion in the next subsection).

\subsection{Relationship between the $K^\#(n,m)$ hierarchy and other nonlinearly dispersive KdV-like equations}
Here, it is worth extending the discussion from Ref.~\refcite{C15} on the connection between the $K^\#(n,m)$ hierarchy introduced in Section~\ref{sec:Ksharp} above and previous models from the literature. For example, in a classic paper from 1993, Rosenau \& Hyman\cite{Rosenau1993} introduced a simple model of a nonlinearly dispersive set of KdV-like equations:
\begin{equation}\label{eq:K}
	K(n,m):\quad u_t + (u^n)_x + (u^m)_{xxx} = 0.
\end{equation}
This set of equations initiated the study of \emph{compactons}, {\it i.e.}, solitary waves of compact support (equivalently, ``finite wavelength''), which has led to an explosion of research on the subject.\cite{Rosenau1993,Rosenau1997,Rosenau1999,Rosenau2000,Rosenau2005,Rosenau2006,Rus2009,Rosenau2014} 
For example, in the case of $n=m=2$ [$K(2,2)$], the exact solution for a compacton is\cite{Rosenau1993}
\begin{equation}\label{eq:K22_compact}
	u(x,t) = 
\begin{cases}
0, &\quad-\infty < x-ct \le -2\pi,\\
\tfrac{4}{3}c\cos^2\left[\tfrac{1}{4}(x-ct)\right], &\quad-2\pi < x-ct < +2\pi,\\
0, &\quad +2\pi \le x-ct \le +\infty.
\end{cases}
\end{equation}
Clearly, $u(x,t)$ is nonzero only on a finite (moving) interval $|x-ct|< 2\pi$. Additionally, $u(x,t)$ is solely a function of the moving-frame coordinate $x-ct$, where $c$ is the speed of the compacton. Finally, just as for the KdV ``$\sech^2$'' soliton,\cite{Drazin1989} the compacton's amplitude, $4c/3$, depends on its speed, $c$.

Unfortunately, since the $K(n,m)$ family of PDEs is based on an {\it ad-hoc} modification of the KdV equation, it does not preserve KdV's Hamiltonian structure. Cooper {\it et al.}\cite{Cooper1993} generalized Rosenau \& Hyman's model\cite{Rosenau1993} to 
\begin{equation}\label{eq:Ks}
	K^*(l,p):\quad u_t + u^{l-2}u_x + \underbrace{u^p u_{xxx} + 2 pu^{p-1}u_x u_{xx} + \tfrac{1}{2}p(p-1)u^{p-2}(u_x)^3} = 0.
\end{equation}
The $K^*(l,p)$ generalization of the $K(n,m)$ equations restores Hamiltonian structure to the model. Equation~\eqref{eq:Ks} extremizes the action generated by the Lagrangian density
\begin{equation}\label{eq:Lagr_Ks}
\mathcal{L}_{K^*}(\varphi;x,t) = \frac{1}{2}\varphi_x\varphi_t + \frac{1}{l(l-1)}(\varphi_x)^l - \frac{1}{2}(\varphi_x)^p(\varphi_{xx})^2.
\end{equation}
Just like the $K(n,m)$ equations, the $K^*(l,p)$ equations possess a variety of compact solitary wave solutions.\cite{Cooper1993,Khare1993,Cooper2006} The equivalent of the $K(2,2)$ compacton solution given in Eq.~\eqref{eq:K22_compact} is the following (quite similar) solution to the $K^*(3,1)$ equation:\cite{Cooper1993}
\begin{equation}\label{eq:Ks31_compact}
	u(x,t) = 
\begin{cases}
0, &\quad-\infty < x-ct \le -\sqrt{3}\pi,\\
3c\cos^2\left[\tfrac{1}{\sqrt{12}}(x-ct)\right], &\quad-\sqrt{3}\pi < x-ct < +\sqrt{3}\pi,\\
0, &\quad +\sqrt{3}\pi \le x-ct \le +\infty.
\end{cases}
\end{equation}

Evidently, the difference between the $K(n,m)$ and $K^*(l,p)$ equations is the ``weights'' of the expanded nonlinearly dispersive term. Specifically, notice that the terms grouped by an under-brace in Eq.~\eqref{eq:Ks} are the same as in the expansion of the term $(u^m)_{xxx}$ (with $m=p+1$) in Eq.~\eqref{eq:K} but with different coefficients! Meanwhile, the difference between the Lagrangian densities in Eqs.~\eqref{eq:Lagr_Ks} and \eqref{eq:Lagr} is in how the dispersive term is modified, {\it i.e.}, from $\propto (\varphi_{xx})^2$ for KdV to $\propto (\varphi_{xx})^{m+1}$ for $K^\#(n,m)$ to $\propto (\varphi_x)^p(\varphi_{xx})^2$ for $K^*(l,p)$. This difference has a profound impact on the structure of traveling wave solutions, as we  explore in detail in Section~\ref{sec:pk_tws} below. Specifically, simple explicit compacton solutions of the form given in Eqs.~\eqref{eq:K22_compact} and \eqref{eq:Ks31_compact} are not available for the $K^\#(n,m)$ equations.

More recently, a version of the $K^*(l,p)$ family of equations, which is invariant under the joint action of parity reflection and time reversal (the so-called ${\cal P}{\cal T}$-symmetric extension\cite{Bender1998} of real-valued PDEs\cite{Bender2007}), has been introduced by Bender {\it et al.},\cite{Bender2009} yielding a hierarchy of equations of the form
\begin{equation}\label{eq:Ks_pt}
	K^*_{{\cal P}{\cal T}}(l,p,r):\quad u_t + u^{l-2}u_x - \frac{p}{r-1}[u^{p-1}(u_x)^r]_x + \frac{r}{r-1}[u^p(u_x)^{r-1}]_{xx} = 0,
\end{equation}
where the third parameter $r$ is necessary to ensure ${\cal P}{\cal T}$-invariance. The corresponding Lagrangian is\cite{Bender2009,Assis2010}
\begin{equation}
\mathcal{L}_{K^*_{{\cal P}{\cal T}}}(\varphi;x,t) = \frac{1}{2}\varphi_x\varphi_t + \frac{1}{l(l-1)}(\varphi_x)^l - \frac{1}{(r-1)} (\varphi_x)^p(\varphi_{xx})^r. 
\end{equation}
Evidently, in introducing the ${\cal P}{\cal T}$-symmetric version of the $K^*(l,p)$ equations, and the new parameter $r$, an ``interpolation'' between the original $K^*(l,p)$ equations and the $K^\#(n,m)$ equations is obtained! Specifically, we see that, upon proper rescaling of $\varphi$, $x$ and $t$, the subset of $K^*_{{\cal P}{\cal T}}(l,p,r)$ equations \eqref{eq:Ks_pt} with $p=0$ can be mapped onto the $K^\#(n,m)$ equations \eqref{eq:Ksharp}, for appropriately chosen $l$ and $r$.

\section{Construction of traveling wave solutions}
\label{sec:pk_tws}

A goal of the present work is to discuss some exact results 
regarding solutions of the $K^\#(n,m)$ hierarchy of equations. The plan of attack is to first reduce the PDE~\eqref{eq:Ksharp} to an ordinary differential equation (ODE) and, then, to study the structure of the latter's solutions.

\subsection{Reduction to an ODE and its integration}

Let $\xi = x-ct$ be moving-frame coordinate for some speed $c>0$ ({\it i.e.}, a right-propagating wave). Then, we suppose that the solution to Eq.~\eqref{eq:Ksharp} can be expressed in the form of a \emph{traveling wave}, specifically $u(x,t) = U(\xi)$. Substituting this {\it ansatz} into Eq.~\eqref{eq:Ksharp}, we arrive at the ODE:
\begin{equation}\label{eq:Ksharp_ode}
	-cU' + U^{n}U' + \big[(U')^{m}\big]'' = 0,
\end{equation}
where primes stand for derivatives with respect to $\xi$. 
Noting that $U^{n}U' = (U^{n+1})'/(n+1)$, the ODE~\eqref{eq:Ksharp_ode} can be immediately integrated once with respect to $\xi$ to yield:
\begin{equation}\label{eq:Ksharp_ode2}
	-cU + \frac{1}{n+1}U^{n+1} + \big[(U')^{m}\big]' = C_1,
\end{equation}
where $C_1$ is an arbitrary constant of integration.
Next, we multiply Eq.~\eqref{eq:Ksharp_ode2} by $U'$ and rewrite all terms as complete derivatives:
\begin{equation}\label{eq:Ksharp_ode3}
	-\frac{c}{2}\big(U^2\big)' + \frac{1}{(n+1)(n+2)}\big(U^{n+2}\big)' + \frac{m}{m+1} \big[(U')^{m+1}\big]' = C_1 U'.
\end{equation}
Integrating a second time and rearraning, we arrive at
\begin{equation}\label{eq:Ksharp_ode4}
	(U')^{m+1} = C_2 + \Psi[U;C_1,\kappa,\gamma],
\end{equation}
where
\begin{equation}
	\Psi[U;C_1,\kappa,\gamma] \equiv  C_1 U + \kappa U^2 - \gamma U^{n+2},
\end{equation}
and, for convenience, we have introduced
\begin{equation}
	\kappa\equiv \frac{(m+1)}{2m}c,\qquad \gamma\equiv \frac{(m+1)}{(n+1)(n+2)m}.
\end{equation}

Taking the $(m+1)$st root of both sides of Eq.~\eqref{eq:Ksharp_ode4} and separating variables yields an \emph{exact but implicit} relation between $\xi$ and $U$:
\begin{equation}\label{eq:Ksharp_impl}
	\int \frac{\mathrm{d}U}{\left(C_2 + \Psi[U]\right)^{1/(m+1)}} = \pm\xi + \xi_0,
\end{equation}
where $\xi_0$ is the final integration constant. The ambiguity introduced by the `$\pm$' sign on  the right-hand side of Eq.~\eqref{eq:Ksharp_impl} is due to the multi-valued nature of the $(m+1)$st root; choosing a branch cut, for example, fixes the sign, but both signs are valid. One obvious restriction that must be applied in arriving at Eq.~\eqref{eq:Ksharp_impl} is that
\begin{equation}\label{eq:Pos_Denom}
C_2 + \Psi[U] \ge 0
\end{equation}
for all $U$ in the (currently unspecified) integration interval, for given values of $C_2$, $C_1$, $\kappa$ and $\gamma$. Hence, allowable limits of integration in Eq.~\eqref{eq:Ksharp_impl} can fall between consecutive zeros of $C_2 + \Psi[U]$ and/or $\pm \infty$, as long as the condition in Eq.~\eqref{eq:Pos_Denom} is satisfied.

More generally, solutions to Eq.~\eqref{eq:Ksharp_ode4} correspond to \emph{integral curves} in the $(U,U')$ phase plane.\cite{Davis1962} Certain values $U = U^*$ such that
\begin{equation}\label{eq:Ksharp_ode_equil}
	C_2 + \Psi[U^*;C_1,\kappa,\gamma] = 0 \quad\Leftrightarrow\quad U'|_{U=U^*} = 0
\end{equation}
are called \emph{equilibria}. The study of solutions to Eq.~\eqref{eq:Ksharp_ode4} can thus be reduced to determining integral curves connecting different equilibria in the $(U,U')$ plane. Clearly, equilibria $U^*$ correspond to singularities of the integrand in Eq.~\eqref{eq:Ksharp_impl}, and their nature must be understood in order to carry out the integration. Moreover, integral curves in the phase plane begin or end at equilibrium points (or go to $\pm\infty$), hence the limits of integration on the left-hand side of Eq.~\eqref{eq:Ksharp_impl} are combinations of $U^*$ values (or $\pm\infty$). (Of course, in the present context, we restrict ourselves to solutions such that $|U(\xi)|<\infty$ $\forall\xi$, on physical grounds.) One consequence of this interpretation is that $\xi$ plays the role of ``time of flight'' along an integral curve in the $(U,U')$ phase plane. Hence, if two equilibria $U=U_{1,2}^*$ can be reached along an integral curve in finite ``time,'' the solutions in the $\big(\xi,U(\xi)\big)$ plane are such that $U(\xi) \ne U_{1,2}^*$  only on a \emph{finite} (in other words, \emph{compact}) interval of $\xi$.

Finally, note that Eq.~\eqref{eq:Ksharp_ode} admits ``anti'' solutions propagating to the left ({\it i.e.}, the ODE remains invariant) under the transformation $U\to-U$ and $c\to -c$ \emph{only if} $n+1$ and $m$ are even.

Let us now illustrate the ideas discussed in this subsection in some special cases.

\subsubsection{$C_1 = C_2 = 0$}
\label{sec:c1c20}

Consider the case in which the first two integration constants are forced to vanish, {\it i.e.}, $C_1 = C_2 = 0$. Then, Eq.~\eqref{eq:Ksharp_ode4} becomes
\begin{equation}\label{eq:Ksharp_ode4_C12_0}
	(U')^{m+1} = U^2(\kappa - \gamma U^{n}).
\end{equation}
Clearly, there are two equilibria in this case:
\begin{equation}\label{eq:equil_C12_0}
	U_1^* = 0,\qquad U_2^* = \left(\frac{\kappa}{\gamma}\right)^{1/n} = \big[\tfrac{1}{2}(n+1)(n+2)c\big]^{1/n}.
\end{equation}
Notice that \emph{unless} $C_2 = 0$, then $U_1^*=0$ \emph{cannot} be an equilibrium. We shall return to the case of $C_1\ne0$ in Section~\ref{sec:c1ne0}.

A peakompacton will thus be a traveling wave solution that connects the two equilibria in Eq.~\eqref{eq:equil_C12_0}. To construct our desired traveling wave solution, we now fix one of the limits of integration in Eq.~\eqref{eq:Ksharp_impl} to be an equilibrium point (specifically, it is convenient to set the lower limit to $U_1^*=0$) to obtain:
\begin{equation}\label{eq:Ksharp_impl_C12_0}
	\int_{0}^U \frac{\mathrm{d}\hat{U}}{\left[ \hat{U}^2 \left(\kappa - \gamma \hat{U}^{n}\right)\right]^{1/(m+1)}} = \pm\xi + \xi_0,
\end{equation}
where $\hat{U}$ is the ``dummy'' integration variable.
Upon judicious inspection of the integrand (or using the computer algebra system {\sc Mathematica}), we find that the integral above is one of the integral representations of Gauss' hypergeometric function ${}_2F_1(a,b;c;z)$.\cite{DLMF} With some effort, Eq.~\eqref{eq:Ksharp_impl_C12_0} evaluates to
\begin{equation}\label{eq:pk_sol}
	\left(\frac{m+1}{m-1}\right) \left[\frac{U^{(m-1)/(m+1)}}{\kappa^{1/(m+1)}}\right] 
 	{}_2F_1 \left[\frac{1}{m+1},\frac{m-1}{(m+1)n};1+\frac{m-1}{(m+1)n};\frac{\gamma}{\kappa}U^n\right] = \pm\xi + \xi_0.
\end{equation}

We must ensure that this solution does indeed pass through the two equilibria given in Eq.~\eqref{eq:equil_C12_0}. By construction, the left-hand side of Eq.~\eqref{eq:pk_sol} already goes to $0$ as $U\to U_1^*$, hence we must have that $ \xi \to \mp\xi_0$ on the right-hand side. This clearly means that the $U_1^*$ equilibrium can be reached in \emph{finite} $\xi$! Then, by translation invariance, we are free to require that $U \to U_2^*$ at any $\xi$, in particular at $\xi=0$. Enforcing this last condition,
\begin{equation}\label{eq:xi0}
	\xi_0 = \left(\frac{m+1}{m-1}\right) \left[\frac{(U_2^*)^{(m-1)/(m+1)}}{\kappa^{1/(m+1)}}\right] 
	{}_2F_1\left[\frac{1}{m+1},\frac{m-1}{(m+1) n};1+\frac{m-1}{(m+1) n};1\right],
\end{equation}
where $U_2^* = \big[\tfrac{1}{2}(n+1)(n+2)c\big]^{1/n}$ as given in Eq.~\eqref{eq:equil_C12_0}.
Thus, the complete peakompacton traveling wave solution is 
\begin{equation}\label{eq:peakompacton}
	U(\xi) = 
\begin{cases}
	0, &\quad -\infty < \xi \le -\xi_0,\\
	\text{inverse of Eq.~\eqref{eq:pk_sol}}, &\quad -\xi_0 < \xi < 0,\\
	\big[\tfrac{1}{2}(n+1)(n+2)c\big]^{1/n}, &\quad\xi = 0,\\
	\text{inverse of Eq.~\eqref{eq:pk_sol}}, &\quad 0 < \xi < + \xi,\\
	0, &\quad +\xi_0 \le \xi \le +\infty.
\end{cases}
\end{equation}
Clearly, this peakompacton is \emph{localized} in the ``usual'' sense of solitons, specifically $U(\xi)\to0$ as $|\xi|\to\infty$ and, hence, $u(x,t)\to0$ as $|x|\to\infty$ ($t<\infty$).

The reason we are allowed to piece (or, ``glue'') together the non-zero solution in Eq.~\eqref{eq:pk_sol} with the equilibrium solution $U=U_1^*=0$ at the (finite) points $\xi=\pm\xi_0$ is that, at those points, the Lipschitz condition (see, {\it e.g.}, Chapter 5, \S8 in Ref.~\refcite{Coddington}) required for the \emph{uniqueness} of solutions of an ODE is violated. This violation is also evident from Eq.~\eqref{eq:Ksharp_ode4_C12_0}, in which solving for $U'$ involves \emph{fractional} powers of the right-hand side; fractional roots are generally not Lipschitz functions. Therefore, many solutions to this nonlinear ODE can be constructed, in particular the peakompacton in Eq.~\eqref{eq:peakompacton} (see also the discussion in Refs.~\refcite{Destrade2006,Jordan2012,Saccomandi2004,Destrade2007,Destrade2009}).

\begin{figure}[!h]
	\centerline{\includegraphics[width=0.9\textwidth]{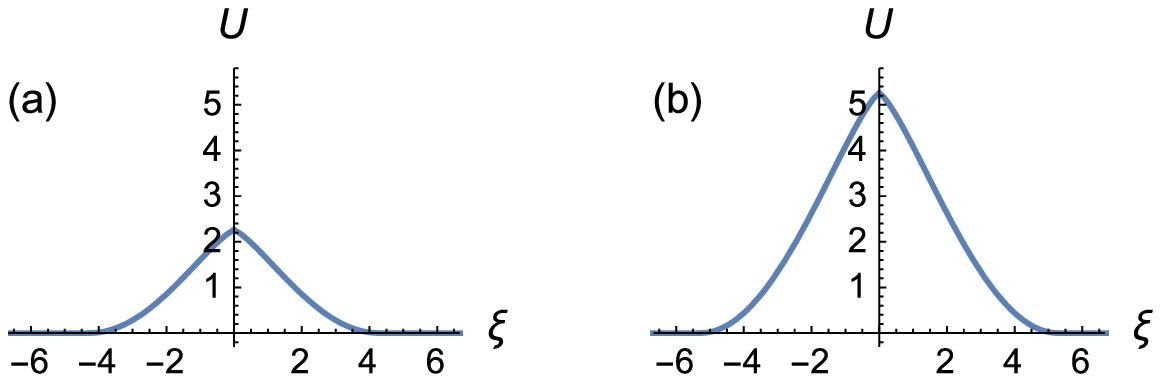}}
	\vspace*{8pt}
	\centerline{\includegraphics[width=0.9\textwidth]{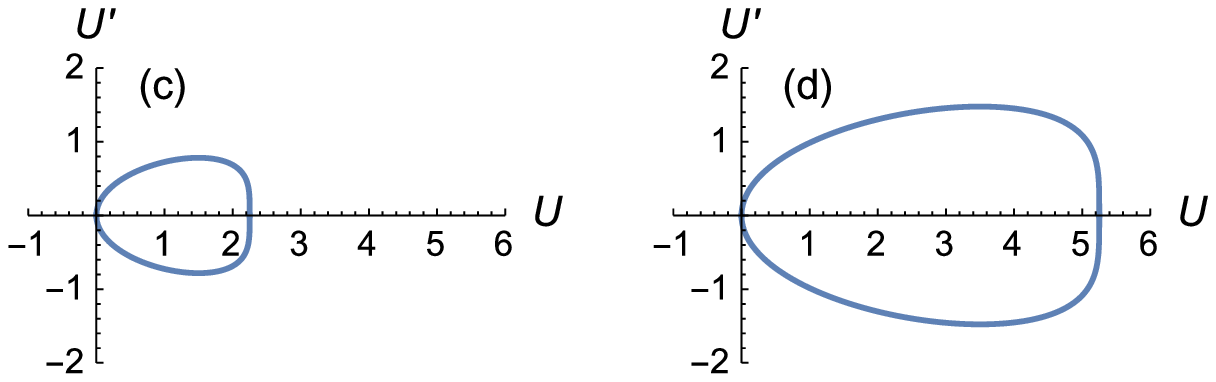}}	
	\vspace*{8pt}
	\caption{The peakompacton solution of the $K^\#(1,3)$ equation for (a) $c=0.75$ ($\Rightarrow\xi_0\approx4.27$) and (b) $c=1.75$ ($\Rightarrow\xi_0\approx5.28$), constructed via Eq.~\eqref{eq:peakompacton}. Panels (c) and (d) show the corresponding phase plane portraits of the peakompactons from panels (a) and (b).}
	\label{fig:pk_illustr}
\end{figure}

From Eq.~\eqref{eq:peakompacton}, it is clear that the amplitude, $U_2^*$, of the peakompacton is solely a function of $n$ (the power of the advective nonlinearity). Meanwhile, $m$ (the power of the dispersive nonlinearity) only affects the width, $2\xi_0$, of the peakompacton. The wave speed $c$ changes both the amplitude and width through $U_2^*$. Figure~\ref{fig:pk_illustr} shows the peakompacton solution of $K^\#(1,3)$ for (a) $c=0.75$ (``subsonic'') and (b) $c=1.75$ (``supersonic''). As is typical for nonlinear waves, such as the KdV ``$\sech^2$'' soliton\cite{Drazin1989} and compactons,\cite{Rosenau1993,Cooper1993} the peakompacton's amplitude is a function of its speed $c$.

\subsubsection{$C_1\ne0$, $C_2=0$}
\label{sec:c1ne0}

In this case, Eq.~\eqref{eq:Ksharp_ode4} becomes
\begin{equation}\label{eq:Ksharp_ode4_C2_0}
	(U')^{m+1} = C_1 U + U^2(\kappa - \gamma U^{n}).
\end{equation}
Now, there can be up to $n+2$ equilibria (zeros of the right-hand side), with one of them being $U_1^*=0$, which clearly persists. By Descartes' rule of signs, however, if $C_1<0$, then the right-hand side of Eq.~\eqref{eq:Ksharp_ode4_C2_0} has at most two further (real) positive roots. Meanwhile, if $C_1>0$, then the right-hand side of Eq.~\eqref{eq:Ksharp_ode4_C2_0} has only one further (real) positive root. For arbitrary $n$, it is not possible to find closed form expressions for these roots.
In the special case of $n=1$, we use the quadratic formula to obtain
\begin{equation}\label{eq:equil_C2_0}
	U_1^* = 0,\qquad U_{2,3}^* = \frac{\kappa \pm \sqrt{4 C_1 \gamma + \kappa^2}}{2 \gamma} \qquad (n=1).
\end{equation}
Similarly, Cardano's formula yields a closed-form expression for $n=2$, however, it is too lengthy to list here.

\begin{figure}[!h]
	\centerline{\includegraphics[width=0.9\textwidth]{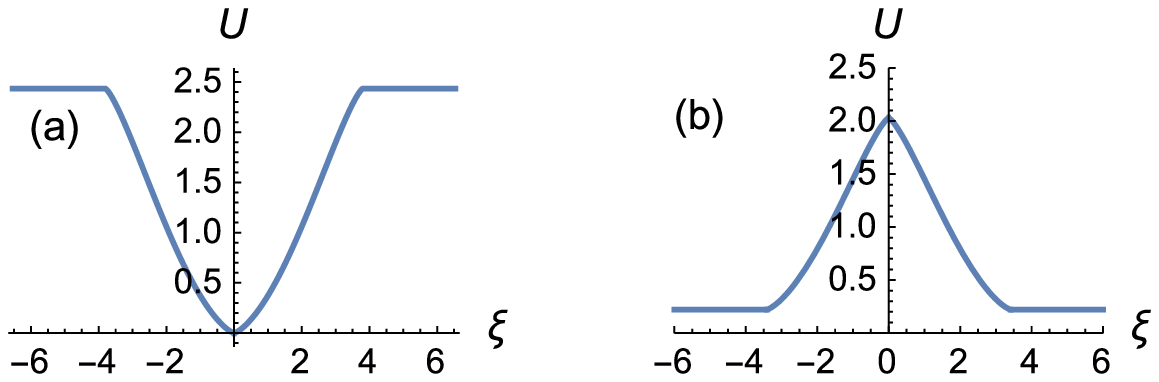}}
	\vspace*{8pt}
	\centerline{\includegraphics[width=0.9\textwidth]{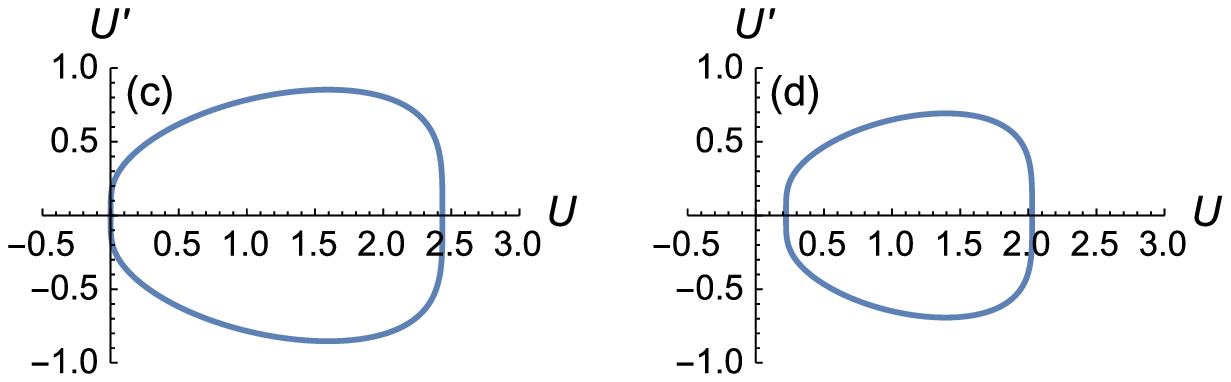}}
	\vspace*{8pt}
	\caption{Nonstandard peakompacton solutions of the $K^\#(1,3)$ equation with $c=0.75$ for (a) $C_1 = 0.1$ ($\Rightarrow U_3^* \approx 2.43$), constructed by piecing together the solution in Eq.~\eqref{eq:pk_Appell} and the equilibrium $U_3^*$ from Eq.~\eqref{eq:equil_C2_0}, and (b) $C_1 = -0.1$ ($\Rightarrow U_2^*\approx 0.22$, $U_3^* \approx 2.02$), constructed by piecing together numerical solutions of Eq.~\eqref{eq:Ksharp_ode4_C2_0} and the equilibrium $U_2^*$ from Eq.~\eqref{eq:equil_C2_0}. Panels (c) and (d) show the corresponding phase plane portraits of the peakompactons from panels (a) and (b).}
	\label{fig:pk_illustr2}
\end{figure}

A qualitative analysis of the phase plane of Eq.~\eqref{eq:Ksharp_ode4_C2_0} [shown in Fig.~\ref{fig:pk_illustr2}(c)] reveals that when $C_1>0$ ({\it i.e.}, when $U_2^* < 0 < U_3^*$), a peakompacton connecting $U_1^*$ and $U_3^*$ can be constructed. However, in this case, the solution turns out to be an \emph{anti}-peakompacton, which travels on a ``level,'' as illustrated in Fig.~\ref{fig:pk_illustr2}(a). A lengthy calculation shows that, although there does not appear to be a closed-form solution (even implicit) for arbitrary $n$, the special case of $n=1$ is amenable to further manipulation, and an implicit solution can be constructed for the wave profile connecting $U_1^*$ and $U_3^*$:
\begin{multline}\label{eq:pk_Appell}
\left(\frac{m+1}{m}\right) \left\{\frac{[(\kappa- \mathfrak{s})U + 2C_1][(\kappa+ \mathfrak{s})U + 2C_1]}{4 C_1^2 [C_1+U (\kappa -\gamma  U)]U} \right\}^{1/(m+1)} U  \\
\times F_1\left(\frac{m}{m+1};\frac{1}{m+1},\frac{1}{m+1};2-\frac{1}{m+1};\frac{2\gamma U}{\kappa +\mathfrak{s}},\frac{2\gamma U}{\kappa -\mathfrak{s}}\right)
= \pm\xi \qquad (n=1),
\end{multline}
where we have set $\mathfrak{s} \equiv \sqrt{4 \gamma C_1 + \kappa^2}$ for convenience, and $F_1(a;b_1,b_2;c;x,y)$ is the Appell hypergeometric function\cite{DLMF2} of two variables. The left-hand side of Eq.~\eqref{eq:pk_Appell} vanishes as $U\to U_1^*=0$, hence we have set $\xi_0 = 0$ to center the traveling wave at $\xi=0$.

In the case when both equilibria $U_{2,3}^* > 0$ [{\it i.e.}, $-\kappa^2/(4\gamma) < C_1 < 0$], we can once again construct a peakompacton connecting $U_2^*$ to $U_3^*$. In this case, we found difficulties adapting the solution given in Eq.~\eqref{eq:pk_Appell}. However, numerical integration (using {\sc Mathematica}'s {\tt NDSolve} subroutine) of Eq.~\eqref{eq:Ksharp_ode4} starting from $U_2^*$ allows us to compute the wave profile, which is now a peakompacton traveling on a ``level,''  as illustrated in Fig.~\ref{fig:pk_illustr2}(b).

\subsubsection{$C_1 = 0$, $C_2 \ne 0$}
\label{sec:c2ne0}

In this case, Eq.~\eqref{eq:Ksharp_ode4} becomes
\begin{equation}\label{eq:Ksharp_ode4_C1_0}
	(U')^{m+1} = C_2 + U^2(\kappa - \gamma U^{n}).
\end{equation}
Once again, it is not possible to determine the equilibrium points of this ODE for arbitrary $n\ge1$. Now, even the case of $n=1$ is non-trivial due to the resulting cubic polynomial on the right-hand side of Eq.~\eqref{eq:Ksharp_ode4_C2_0}.

The most pernicious feature of having $C_1 = 0$ and $C_2 \ne 0$ is that $U=0$ is no longer an equilibrium, which was always the case in Sections~\ref{sec:c1c20} and \ref{sec:c1ne0}. By Descartes' rule of signs, if $C_2<0$, then the right-hand side of Eq.~\eqref{eq:Ksharp_ode4_C1_0} has only a single positive root. Therefore, for $C_2<0$, a peakompacton solution does not exist. On the other hand, it is easy to see that the case of $C_2>0$ is qualitatively similar to the case of $C_2=0$ and $C_1<0$ from Section~\ref{sec:c1ne0}, so we will not dwell on it further.

\subsubsection{Other cases}

It is certainly conceivable that even more traveling wave constructions, for special values of the integration constants $C_1$, $C_2$ and the exponents $n$, $m$, are possible. We do not claim the discussion above exhausts all possibilities. However, the cases considered above do illustrate a breadth of possible ``exotic'' traveling wave solutions to the $K^\#(n,m)$ hierarchy of equations.

For example, periodic solutions of arbitrary spatial period $\ell \ge 2\xi_0$, where $[-\xi_0,+\xi_0]$ is the interval of compact support of the solutions above, can always be constructed from identical peakompactons ({\it i.e.}, same $n$, $m$ and $c$). This superposition property follows immediately from the compactness (finite wavelength) of the peakompacton traveling waveforms of the $K^\#(n,m)$ hierarchy of equations.

\subsection{Regarding derivative discontinuities}

Although the class of solutions derived in Section~\ref{sec:c1c20} ($C_1=C_2=0$) is implicit, it is still possible to obtain results regarding the derivatives $U'(\xi)$ and $U''(\xi)$ of the traveling wave profile. The key fact is that as $\xi\to\pm\xi_0$, $U\to (U_1^*)^+$ and as $\xi\to0$, $U\to (U_2^*)^-$. Then, from Eq.~\eqref{eq:Ksharp_ode4_C12_0}, we obtain 
\begin{equation}
U'(\xi) \to 
\begin{cases}
	0, &\quad \xi\to0,\\
	0, &\quad \xi\to\pm\xi_0.
\end{cases}
\end{equation}
Therefore, $U'(\xi)$, like $U(\xi)$ itself, is continuous at $\xi=\{0, \pm\xi_0\}$ and, hence, for all $\xi \in \mathbb{R}$.

Then, by implicit differentiation, Eq.~\eqref{eq:Ksharp_ode4_C12_0} also allows us to compute
\begin{equation}
	U''(\xi) = \frac{\left\{2\kappa - (n+2)\gamma [U(\xi)]^{n}\right\}U(\xi)}{(m+1)\left([U(\xi)]^2\left\{\kappa - \gamma [U(\xi)]^{n}\right\}\right)^{(m-1)/(m+1)}}.
\end{equation}
Once again, considering the limits $\xi\to\pm\xi_0$ [$U\to (U_1^*)^+$] and $\xi\to0$ [$U\to (U_2^*)^-$], a preliminary analysis of discontinuities of $U''(\xi)$ can be performed; see \S4 of Ref.~\refcite{C15}.
To summarize:
\begin{equation}\label{eq:Upp_limits}
|U''(\xi)| \to 
\begin{cases}
	0, &\quad \xi\to\pm\xi_0\quad (0<m<3),\\
	\sqrt{c/6}, &\quad \xi\to\pm\xi_0\quad (m=3),\\
	\infty,	&\quad \xi\to\pm\xi_0\quad (m>3),\\
	\infty, &\quad \xi\to0\quad (\forall m>0).\\
\end{cases}
\end{equation}

Jump discontinuities of finite size in the higher derivatives of a function [$U(\xi)$ in the present context] are termed \emph{mild discontinuities}.\cite{Coleman1967,Wei2013} More severe discontinuities in the higher derivatives ({\it i.e.}, the derivatives approach $\pm\infty$ or are undefined)  are also possible according to Eq.~\eqref{eq:Upp_limits}. Specifically, higher derivative discontinuities at the \emph{crest} of a wave play an important role in the classification of ``exotic'' traveling wave solutions of nonlinear evolution equations. Another classical nonlinearly dispersive equation, which we have not mentioned so far, is the Camassa--Holm (CH) equation:\cite{Camassa1993,Camassa1994,Boyd1997,Camassa2003}
\begin{equation}\label{eq:CH}
u_t + 3uu_x -2 u_xu_{xx} + uu_{xxx} = u_{xxt},
\end{equation}
which has the following traveling wave solution:
\begin{equation}\label{eq:peakon}
u(x,t) = U(\xi) = c \,\mathrm{e}^{-|\xi|},\qquad \xi \equiv x-ct.
\end{equation}
The solution given in Eq.~\eqref{eq:peakon} is termed a \emph{peakon} because $U'(\xi)$ suffers a finite jump (mild discontinuity) at $\xi=0$, while $U''(\xi)$ is undefined there (due to the presence of a Dirac $\delta$-function in $U''$). Although Eq.~\eqref{eq:CH} is not exactly a nonlinearly dispersive KdV-like equation due to the mixed-derivative term, the CH equation also possesses a rich geometrical structure,\cite{Holm1998,Holm2009} including being \emph{bi-Hamiltonian}\cite{Camassa1993,Camassa1994,Camassa2003} and related to umbilic geodesics on surfaces.\cite{Alber1995} [Interestingly, a change of signs in Eq.~\eqref{eq:CH} can lead to an integrable bi-Hamiltonian equation that admits compacton solutions.\cite{Olver1996}]

\begin{figure}[!h]
	\centerline{\includegraphics[width=0.9\textwidth]{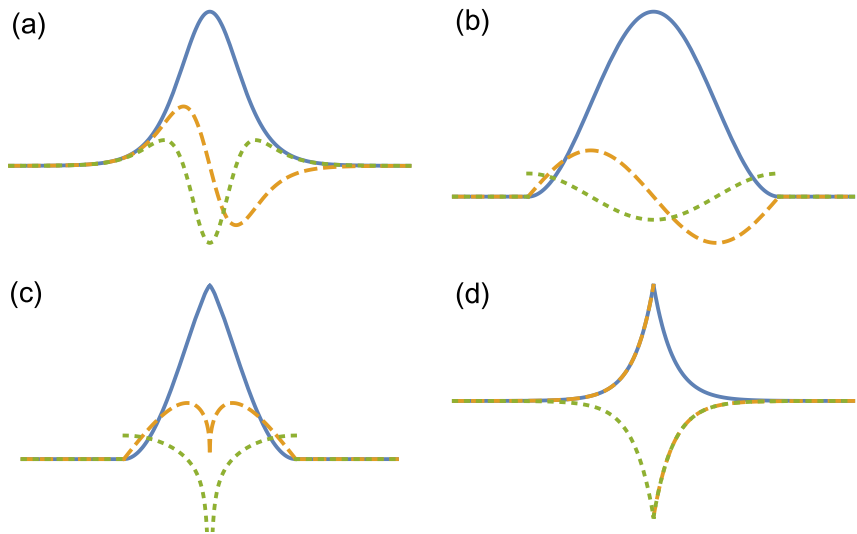}}
	\vspace*{8pt}
	\caption{Illustrations of the traveling wave solutions of four nonlinearly dispersive evolution equations: (a) KdV, (b) $K(2,2)$, (c) $K^\#(1,3)$, and (d) CH. While the horizontal scales are the same in all four panels, the vertical scales have been adjusted for clarity; $c=1$ in all panels. Axes are not shown in order to clearly highlight only \emph{qualitative} features. In each panel, solid curves represent the traveling wave profile $U(\xi)$, dashed curves represent $U'(\xi)$, and dotted curves represent $U''(\xi)$. Note that in panel (c) the vertical axis does not show the full range and $U''(\xi)\to -\infty$, in fact. In panel (d), the Dirac $\delta$-function in $U''(\xi)$ cannot be visualized, of course.}
	\label{fig:twss}
\end{figure}

Figure~\ref{fig:twss} qualitatively illustrates various solitary wave solutions of dispersive evolution equations and their derivatives. The KdV ``$\sech^2$'' soliton\cite{Drazin1989} (a) is $C^\infty(\mathbb{R})$ (infinitely continuously differentiable). The $K(2,2)$ profile\cite{Rosenau1993} (b) is $C^1(\mathbb{R})$ (once continuously differentiable), since $U''$ suffers jumps at the edges of the compact support. The $K^\#(1,3)$ profile\cite{C15} (in the case of $C_1=C_2=0$ from Section~\ref{sec:c1c20}) (c) is $C^1(\mathbb{R})$ as well, with $U''$ suffering jumps at the edges of the compact support and $U''$ blowing up at $\xi=0$. Finally, the CH profile\cite{Camassa1993} (d) is only $C^0(\mathbb{R})$ because $U'$ suffers a jump at $\xi=0$, and $U''$ is undefined there. The important observation here, in particular as far as the terminology is concerned, is that the classical (KdV) soliton is infinitely smooth, while compactons exhibit mild discontinuities at the edges of their compact support. Meanwhile, peakons suffer mild discontinuities at the crest of the wave. Peakompactons exhibit both the features of peakons and compactons, hence the {\it portmanteau} ``peakompacton.''

In passing, we also make note of a \emph{completely integrable}\cite{Hunter1994} evolution equation (unlike the, presumably, non-integrable compacton and peakompacton hierarchies discussed above) that admits piecewise linear solutions, namely the Hunter--Saxton\cite{Hunter1991} equation:
\begin{equation}\label{eq:HS}
(u_t + uu_x)_{xx} = \tfrac{1}{2}\big(u_x^2\big)_x,
\end{equation}
which is an asymptotic model of the dynamics of nematic liquid crystals.
Piecewise linear solutions (in the form of a ``hat function'') of Eq.~\eqref{eq:HS} of compact support and suffering only mild discontinuities at the crest and the boundaries of their support can be constructed. Although such solutions are conceptually related to our discussion, Eq.~\eqref{eq:HS} is fundamentally of a different type than the KdV-like nonlinearly dispersive equations that are the subject of this work. A conceptually closer type of nonlinearly dispersive evolution equation is the Harry Dym equation:\cite{Kruskal1975}
\begin{equation}
	u_t = u^3 u_{xxx}.
\end{equation}
This equation, like the CH equation, admits peakon solutions and is also {completely} integrable by the inverse scattering transform, which brings about connections to the KdV equation.\cite{Hereman1989}

Finally, we acknowledge that the possibility of infinite second derivatives technically means that the solutions considered herein are (in a sense) \emph{weak solutions}, {\it i.e.}, they do not possess as many continuous derivatives as are present in the governing ODE. There are a number of mathematical issues that must be elucidated for such \emph{pseudo-classical}\cite{Li1997,LiOlver1997,LiOlver1998} (or \emph{singular}\cite{Geyer2016}) solutions. For further details, the reader is referred to the comprehensive mathematical discussion by Yi \& Olver.\cite{LiOlver1997,LiOlver1998} From the physical point of view, it suffices to note that compactons and peakompactons do satisfy their respective ODEs in a proper sense because, even if $U$ has discontinuous (or undefined) higher derivatives, it is the \emph{products} of $U$ and its derivatives that appear in the original ODE, {\it e.g.}, Eq.~\eqref{eq:Ksharp_ode}. Therefore, it suffices that those terms have finite limits as $\xi\to\{0,\pm\xi_0\}$, so that the ODE itself does not suffer a jump discontinuity at those points (see also the discussion in Ref.~\refcite{Bender2009}).

\subsection{Explicit variational approximations}
\label{sec:va}

In the previous subsections, we discussed the structure of the \emph{exact} traveling wave reduction of the $K^\#(n,m)$ equations. Clearly, the structure of the associated ODE is nontrivial. Furthermore, the exact solutions we found were implicit and involved hypergeometric special functions. In this subsection, we would like to ask the question: can \emph{explicit approximations} to the traveling wave solutions of the $K^\#(n,m)$ hierarchy of equations given in Eq.~\eqref{eq:Ksharp}, {\it i.e.}, approximate solutions of the ODE in Eq.~\eqref{eq:Ksharp_ode}, be obtained?

To answer this question, we appeal to the so-called method of \emph{variational approximation}\cite{Sugiyama1979,Anderson1983,Rice1983,Campbell1983} (see also Refs.~\refcite{Kaup2007,Chong2012} for discussion of the method's accuracy). The idea of the variational approximation is that, for a given properly parametrized explicit functional form of the traveling wave solution $U(\xi)$, from which $u(x,t)$ and $\varphi(x,t)$ follow of course, the variational structure of the governing PDE provides a natural way to determine ``optimal'' choices of the free parameters in the parametrized approximation. This technique is best illustrated through an example.

Ideally, a parametrized explicit functional form, or {\it ansatz}, for peakompactons would itself also be compact. We could introduce a trial function of the form
\begin{equation}\label{eq:anz}
U(\xi) \approx \tilde{U}(\xi;t) = \begin{cases}
\mathfrak{A}(t) \mathrm{e}^{\mathfrak{b}(t)/(|\xi|-\xi_\mathrm{c})}, &|\xi| < \xi_\mathrm{c},\\
0, &|\xi| \ge \xi_\mathrm{c},\end{cases}
\end{equation}
where $\mathfrak{A}(>0)$ and $\mathfrak{b}(>0)$ are to be determined as part of the procedure, $\xi \equiv x-ct$ is the moving-frame coordinate as before, and $\xi_c$ is the half-length of the compact support [{\it e.g.}, $\xi_0$ as given in Eq.~\eqref{eq:xi0} for the peakompactons constructed in Section~\ref{sec:c1c20}]. The key idea is to now compute $\varphi(x,t)$ on the basis of Eq.~\eqref{eq:anz} and, then, substitute the expression into the Lagrangian {density} $\mathcal{L}(\varphi;x,t)$ from Eq.~\eqref{eq:Lagr}. The next step is to compute $\int^{+\infty}_{-\infty} \mathrm{d}x\,\mathcal{L}$ to obtain the Lagrangian itself. Because the calculated Lagrangian is based on an {\it ansatz} and is no longer a function of $x$, we term it the \emph{coarse-grained Lagrangian}\cite{Christov2008} $\mathbb{L}(\mathfrak{A},\mathfrak{b};t)$. Now, $\mathbb{L}$ generates an action functional, which we must require to be stationary with respect to variations of $\mathfrak{A}$ and $\mathfrak{b}$. Hence, two coupled ODEs can be obtained, which determine the ``optimal'' values of the \emph{a priori} undetermined parameters in the trial function $\tilde{U}(\xi;t)$. 

\begin{figure}
	\centerline{\includegraphics[width=0.6\textwidth]{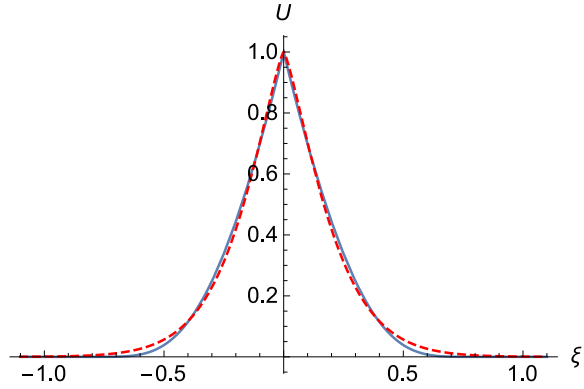}}
	\vspace*{8pt}
	\caption{Illustration of the differences between the analytically intractable but compact {\it ansatz} given in Eq.~\eqref{eq:anz} (solid curve) and the  analytically tractable but non-compact {\it ansatz} given in Eq.~\eqref{eq:anz2} (dashed curve); $\mathfrak{A} = 25.79$, $\mathfrak{b}=3.25$, $\xi_\mathrm{c}=1$ and $\mathcal{A} = 1$, $\beta=4.5$, $\eta = 0.65$. Evidently, these two {\it ans\"atze} can be made to agree quite well for appropriately chosen (for the purposes of this figure, ``by eye'') parameters.}
	\label{fig:anz_diff}
\end{figure}

Unfortunately, however, when using Eq.~\eqref{eq:anz} as the {\it ansatz}, the integration over $x$ cannot be performed in terms of elementary functions. An alternative parametrized functional form is required. One way forward is to relax the requirement that the {\it ansatz} be a compact function of $\xi$, and to use the \emph{post-Gaussian} trial function:\cite{Cooper1993,Bender2009,Cooper2001}
\begin{equation}\label{eq:anz2}
	U(\xi) \approx \tilde{U}(\xi) = {\cal A} \exp\left\{-\beta|\xi|^{2\eta}\right\},
\end{equation}
where ${\cal A}(>0)$, $\beta(>0)$ and $\eta(>0)$ are, {\it a priori}, unknown and are \emph{no longer} functions of time. The difference between the {\it ans\"atze} in Eqs.~\eqref{eq:anz} and \eqref{eq:anz2} is illustrated in Fig.~\ref{fig:anz_diff}.

Integrating Eq.~\eqref{eq:anz2} with respect to $\xi$, we find the functional form of the field $\varphi(x,t) = \Phi(\xi)$ associated to $u(x,t)$ under the traveling wave assumption:
\begin{equation}
	\tilde\Phi(\xi) = -\frac{\mathcal{A} \xi}{2 \eta \left(\beta  |\xi|^{2 \eta }\right)^{1/(2 \eta)}} \Gamma \left(\frac{1}{2 \eta },\beta  |\xi|^{2 \eta }\right),
\end{equation}
where $\Gamma(a,x)\equiv \int_x^{+\infty}\mathrm{d}\zeta\, \zeta^{a-1} \mathrm{e}^{-\zeta}$ is the \emph{incomplete Gamma function}.\cite{IG} To compute $\int^{+\infty}_{-\infty} \mathrm{d}x\,\mathcal{L}$ based on the Lagrangian density in Eq.~\eqref{eq:Lagr}, we note the following chain rules apply for the chosen {\it ansatz}:
\begin{equation}\label{eq:chain_rules}
	\varphi_x(x,y) = \tilde\Phi'(\xi) \xi_x = \tilde U(\xi),\quad \varphi_{xx}(x,t) = \tilde U'(\xi),\quad \varphi_t(x,t) = \tilde\Phi'(\xi)\xi_t = -c\tilde U(\xi).
\end{equation}
Then, substituting the expressions from Eq.~\eqref{eq:chain_rules} into $\int^{+\infty}_{-\infty} \mathrm{d}x\,\mathcal{L}$, using the fact that $\mathrm{d}x = \mathrm{d}\xi$ and appealing to symmetry to rewrite $\int^{+\infty}_{-\infty}\mathrm{d}\xi\, [\cdots] = 2\int^{+\infty}_{0}\mathrm{d}\xi\, [\cdots]$, we obtain
\begin{equation}
	\mathbb{L} = 2\int_0^{+\infty} \mathrm{d}\xi \left[ -\frac{c}{2}\tilde U^2 + \frac{1}{(n+2)(n+1)} \tilde U^{n+2} - \frac{1}{m+1}\big(\tilde U'\big)^{m+1}\right].
\end{equation}
Upon evaluating the integrals using $\tilde U$ from Eq.~\eqref{eq:anz2}, the coarse-grained Lagrangian takes the form
\begin{multline}\label{eq:Lagr_VA}
	\mathbb{L}(\mathcal{A},\beta,\eta;n,m,c) = \Gamma \left(1+\frac{1}{2 \eta }\right) \left[ \frac{2 \beta  \mathcal{A}^{n+2}}{(n+1)[\beta (n+2)]^{1+1/(2\eta)}} - \frac{c\mathcal{A}^{2}}{(2\beta)^{1/(2\eta)}} \right] \\
	+ (-1)^m \frac{(2\mathcal{A})^{m+1} (\beta  \eta )^m \Gamma \left[\left(1-\frac{1}{2 \eta}\right)m+1\right] }{(m+1)^2[\beta  (m+1)]^{[1-1/(2\eta)] m}}.
\end{multline}
This expression for $\mathbb{L}$ is valid only for $\eta$ such that $m<2 \eta  (m+1)$, otherwise the integral defining $\mathbb{L}$ does not converge.
Now, we extremize the function of three variables (all of which are independent of time) given in Eq.~\eqref{eq:Lagr_VA}, namely we require that the following simultaneous equations hold:
\begin{equation}\label{eq:L_extrem}
\left\{\frac{\partial \mathbb{L}}{\partial \mathcal{A}} = 0,\quad \frac{\partial \mathbb{L}}{\partial \beta} = 0,\quad \frac{\partial \mathbb{L}}{\partial \eta} = 0\right\}.
\end{equation}
The resulting solutions for ${\cal A}$, $\beta$ and $\eta$ in terms of each other and $n$, $m$ and $c$ are lengthy (see also the discussion in \S IV in Ref.~\refcite{Cooper1993}). To illustrate the variational approximation, however, let us restrict to the two featured equations from Section~\ref{sec:motive}, namely $K^\#(1,3)$ and $K^\#(2,3)$. Furthermore, let us fix the wave speed to be $c=0.75$ (a ``subsonic'' peakompacton). Then, the three conditions in Eq.~\eqref{eq:L_extrem} yield
\begin{equation}\label{eq:Abh_123}
(\mathcal{A},\beta,\eta) \approx
\begin{cases}
(2.21423,0.310367,0.822982), &\quad (n,m,c) = (1,3,0.75),\\
(2.09741,0.428176,0.774748), &\quad (n,m,c) = (2,3,0.75).
\end{cases}
\end{equation}
For completeness, note that the system in Eq.~\eqref{eq:L_extrem} is a non-trivial set of transcendental equations coupling ${\cal A}$, $\beta$ and $\eta$. This system is amenable by numerical methods only. In practice, it is useful and possible to first solve $\partial\mathbb{L}/\partial\beta =0$ explicitly for $\beta$ in terms of $\mathcal{A}$ and $\eta$ (as well as $n$, $m$ and $c$, of course). Then, the latter solution is substituted into the remaining two equations, which can then be solved numerically using {\sc Mathematica}'s {\tt FindRoot} subroutine. 

\begin{figure}[!h]
	\centerline{\includegraphics[width=0.9\textwidth]{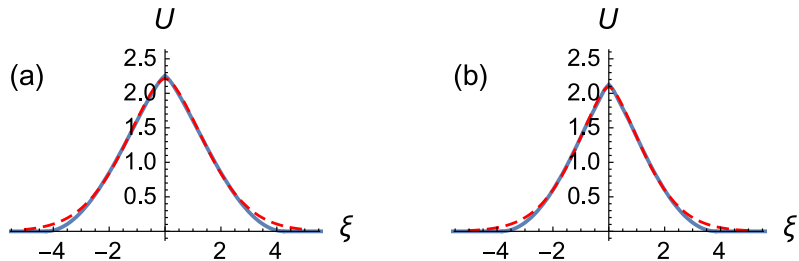}}
	\vspace*{8pt}
	\caption{Comparison between the exact peakompacton solutions [{\it i.e.}, Eq.~\eqref{eq:peakompacton}, solid curves] of the (a) $K^\#(1,3)$ and (b) $K^\#(2,3)$ equations and their variational approximations [{\it i.e.}, Eqs.~\eqref{eq:anz2} and \eqref{eq:Abh_123}, dashed curves].}
	\label{fig:va_compare}
\end{figure}

A comparison between the exact peakompacton solutions and their variational approximations is shown in Fig.~\ref{fig:va_compare}; the agreement is quite good away from $\xi=\pm\xi_0$, where the \emph{non}-compact {\it ansatz} in Eq.~\eqref{eq:anz2} is not expected to be a good approximation. Note that $\max_\xi U(\xi) = U_2^* \equiv \big[\tfrac{1}{2}(n+1)(n+2)c\big]^{1/n}$ [recall Eq.~\eqref{eq:equil_C12_0}], which gives $\max_\xi U(\xi) = 2.25$ for $(n,m,c) = (1,3,0.75)$ and $\max_\xi U(\xi) \approx 2.12132$ for $(n,m,c) = (2,3,0.75)$. On comparing these values to the corresponding $\mathcal{A}$ values given in Eq.~\eqref{eq:Abh_123}, we see that the variational approximation is accurate to $\approx 1.5\%$. 

Finally, once the optimal values of ${\cal A}$, $\beta$ and $\eta$ have been computed for given $n$, $m$ and $c$, the total energy, wave mass and wave momentum can easily be found from Eqs.~\eqref{eq:HMP} based on the {\it ansatz} given in Eq.~\eqref{eq:anz2}:
\begin{subequations}
\begin{align}
H &= \frac{(-1)^{m+1}(2\mathcal{A})^{m+1} (\beta \eta)^m \Gamma\left[\left(1-\frac{1}{2\eta}\right) m + 1\right]}{(m+1)^2[\beta (m+1)]^{[1-1/(2\eta)] m}} -\frac{2 \beta  \mathcal{A}^{n+2} \Gamma \left(1+\frac{1}{2 \eta }\right) }{(n+1)[\beta (n+2)]^{1+1/(2\eta)}},\\
M &= \frac{\mathcal{A}}{\beta^{1/(2\eta)}} \Gamma\left(1+\frac{1}{2\eta}\right),\\[3mm]
P &= - \frac{\mathcal{A}^2}{(2\beta)^{1/(2\eta)}} \Gamma\left(1+\frac{1}{2\eta}\right).
\end{align}
\end{subequations}
As before, the expression for $H$ is valid only for $\eta$ such that $m<2 \eta  (m+1)$, otherwise the integral defining $H$ does not converge.

\section{Numerical solution of peakompacton equations}

Now, we turn to the numerical solution of the $K^\#(n,m)$ equations given in Eq.~\eqref{eq:Ksharp}. Our purpose in obtaining numerical solutions to the ``full'' PDE is to simulate and shed light on the interactions of multiple peakompactons.

Previously Pad\'e approximation methods,\cite{Rus2007b,Mihaila2010a,Mihaila2010b,Cardenas2011} finite difference\cite{Vlieg1971,Frutos1995} and finite element methods,\cite{Ismail1998} particle methods,\cite{Chertock2001} the discontinuous Galerkin method,\cite{Levy2004} and the method of lines\cite{Saucez2004} have all been used successfully to simulate the evolution of compactons and related KdV-like equations. Here, we use simple finite-difference methods\cite{Strikwerda2004,Lynch2005} (reminiscent of those used in the original compacton simulations\cite{Rosenau1993}) to make the simulation of peakompactons accessible to students and those who are only beginners in numerical mathematics. However, we do warn the reader that there are many intricate details regarding the high-resolution simulation of compacton collisions that have been discussed in detail in the literature,\cite{Rus2007a,Rus2008,Garralon2013} and the reader should be aware of them when attempting to simulate peakompacton collisions.

\subsection{A leap-frog scheme with filtering}
\label{sec:leapfrog}

We begin the construction of our numerical scheme by first \emph{semi-discretizing} Eq.~\eqref{eq:Ksharp} using central differences in space:
\begin{equation}\label{eq:sd_scheme}
	\frac{\mathrm{d}u_j}{\mathrm{d}t} = \underbrace{- \left\{ \frac{2}{3}D_0\Big[\tfrac{1}{(n+1)}(u_j)^{n+1}\Big] + \frac{1}{3} (u_j)^{n}D_0[u_j] \right\} + D^2\left[(D_0[u_j])^{m}\right]}_{\equiv \mathcal{L}_h[u_j]},
\end{equation}
where $u_j(t) \approx u(x_j,t)$ is the semi-discrete approximate solution defined on the $N$-point grid $x_j = -L + j\Delta x$ with $\Delta x = 2L/(N-1)$ and $j=0,1,\hdots,N-1$. We solve Eq.~\eqref{eq:Ksharp} on the finite domain $[-L,L]$ with periodic boundary conditions, {\it i.e.}, $u(-L,t) = u(L,t)$ $\forall t\ge 0$.

In Eq.~\eqref{eq:sd_scheme}, we have employed the difference operators
\begin{equation}
	D_0[u_j] \equiv \frac{-u_{j+2}+8u_{j+1}-8u_{j-1}+u_{j-2}}{12\Delta x},
\end{equation}
which is the fourth-order-accurate central difference approximation to $\partial u/\partial x$ (see Table 2.6 of Ref.~\refcite{Lynch2005}), and 
\begin{equation}
	D^2[u_j] \equiv \frac{-u_{j+2} + 16 u_{j+1} - 30 u_j + 16 u_{j-1} - u_{j-2}}{12(\Delta x)^2},
\end{equation}
which is the fourth-order-accurate central difference approximation to $\partial^2 u/\partial x^2$ (again, see Table 2.6 of Ref.~\refcite{Lynch2005}). Here, one can ``plug-in'' one's favorite higher-order central difference formul\ae\ for the first and second derivative as well. Additionally, we have split the nonlinear advective term $u^nu_x$ into the equivalent form $\frac{2}{3}[\tfrac{1}{(n+1)}u^{n+1}]_x + \frac{1}{3}u^n u_x$ for improved conservation of the quadratic invariant of $u$, {\it i.e.}, the total wave momentum $P$ given in Eq.~\eqref{eq:P_invar}. The linear invariant of $u$, {\it i.e.}, the total wave mass $M$ given in Eq.~\eqref{eq:M_invar} is automatically conserved by virtue of using spatial central difference operators in Eq.~\eqref{eq:sd_scheme}.

Next, we must choose an appropriate time discretization for the semi-discrete system in Eq.~\eqref{eq:sd_scheme}. Following Hyman,\cite{Hyman1979} we discretize the time derivative using an \emph{explicit} three-level predictor--corrector method (of second-order accuracy) known as ``leap-frog 2--3'':
\begin{subequations}\label{eq:leapfrog}\begin{align}
	u^*_j &= u^k_j + 2 \Delta t \mathcal{L}_h[u^{k+1}_j]\qquad &(\text{predictor}), \\\
	u^{k+2}_j &= \tfrac{1}{5}\left\{ u^{k}_j + 4u^{k+1}_j + 4\Delta t\mathcal{L}_h[u^{k+1}_j] + 2\Delta t \mathcal{L}_h[u^{*}_j] \right\}\qquad &(\text{corrector}),
\end{align}\end{subequations}
where $\Delta t$ is a fixed time-step size, $u_j^k \approx u(x_j,t^k)$ and $t^k = k\Delta t$.
This time-integration scheme has desirable stability properties in that it includes portions of the imaginary axis for finite $\Delta t$ (unlike the traditional leap-frog scheme).\cite{Hyman1979} Since Eq.~\eqref{eq:Ksharp} includes dispersive terms, we expect that the discrete spatial operator $\mathcal{L}_h$ in Eq.~\eqref{eq:sd_scheme} to have imaginary and complex eigenvalues, hence our choice of the leap-frog 2--3 scheme for the time discretization.

As mentioned above, the time-stepping scheme in Eq.~\eqref{eq:leapfrog} is a \emph{three-level scheme}. To initialize it, we use the FTCS (forward-time central-space) scheme, {\it i.e.}, we perform an initialization time step by discretizing the left-hand side of Eq.~\eqref{eq:sd_scheme} using the forward Euler scheme: 
\begin{equation}
	\frac{\mathrm{d}u_j}{\mathrm{d}t} \approx \frac{u^1_j - u^0_j}{\Delta t} = \mathcal{L}_h[u^0_j], 
\end{equation}
where $u^0_j$ is the initial condition evaluated on the computational grid.

\begin{figure}
	\centerline{\includegraphics[width=0.6\textwidth]{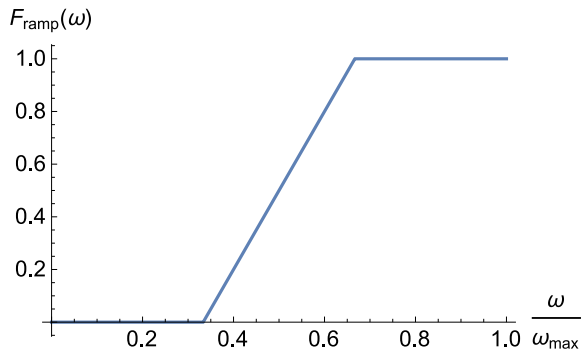}}
	\vspace*{8pt}
	\caption{Fourier space representation $F_\mathrm{ramp}(\omega)$ of the ``ramp'' spectral filter used to define the artificial viscosity in Eq.~\eqref{eq:articifial_visc} for the numerical scheme.}
	\label{fig:ramp}
\end{figure}

Finally, following Cooper {\it et al.},\cite{Cooper2001} we stabilize the numerical method with low-order filtered artificial viscosity by modifying the discrete spatial differential operator as follows
\begin{equation}\label{eq:articifial_visc}
	\mathcal{L}_h[u_j] \mapsto \mathcal{L}_h[u_j] + \text{\tt ramp\_spectral\_filter}\big\{ \nu \Delta x D^2[u_j]\big\},
\end{equation}
where $\nu \Delta x$ is the artificial viscosity. Since the artificial viscosity scales with $\Delta x$, it vanishes as $\Delta x\to0$, hence this modification does not affect the scheme's consistency.
Here, $\text{\tt ramp\_spectral\_filter}\{\,\cdot\,\}$ is a \emph{high-pass} filter that takes the fast Fourier transform (in $x$) of its argument, then multiplies by $0$ the lowest third of the Fourier modes, multiplies by $1$ the highest third of the Fourier modes, and ``ramps'' the  middle third (as illustrated in Fig.~\ref{fig:ramp}), before fast inverting back to real space. Artificial viscosity helps improve the stability of the numerical method by damping high-frequency numerical (aliasing) errors arising from the higher-derivative discontinuities of peakompactons. The spectral filtering step no longer allows us to \emph{prove} that $M$ is conserved automatically, however, we observe in all simulations that $M$ is conserved to better than $1\%$. 

Finally, we expect that such an explicit scheme will be stable only under a Courant--Friedrichs--Lewy (CFL) condition (see, {\it e.g.}, \S1.6 in Ref.~\refcite{Strikwerda2004}):
\begin{equation}\label{eq:CFL}
	\Delta t = C_\mathrm{CFL}(\Delta x)^3,
\end{equation}
for some CFL number $C_\mathrm{CFL} \le 1$. The reason $\Delta x$ is taken to the third power is that the highest spatial derivative in the PDE \eqref{eq:Ksharp} is of third order.


\subsection{Example: an overtaking collision in $K^\#(1,3)$}

In the spirit of the work of Zabusky \& Kruskal,\cite{Zabusky1965} we would like to now establish whether peakompactons can ``survive'' overtaking collisions, and, if so, whether this type of collision is elastic. To this end, we restrict to the case of $n=1$ and $m=3$, {\it i.e.}, the $K^\#(1,3)$ equation. We generate an initial condition that consists of two peakompactons of disjoint support and different wave speeds $c_1$ and $c_2$:
\begin{equation}
	u(x,0) = U(\xi-\xi_1;t=0,c=c_1) + U(\xi-\xi_2;t=0,c=c_2),
\end{equation}
where $U(\xi)$ is given by Eq.~\eqref{eq:peakompacton}. The peakompactons comprising the initial condition are shifted by $\xi_1$ and $\xi_2$ with respect to the origin of the coordinate system to ensure their supports do not overlap and to give them enough distance to propagate before reaching the (periodic) downstream boundary at $x=L$. Then, the full initial--boundary--value problem is solved using the finite-difference scheme constructed in Section~\ref{sec:leapfrog}.

\begin{figure}[b]
	\centerline{\includegraphics[width=\textwidth]{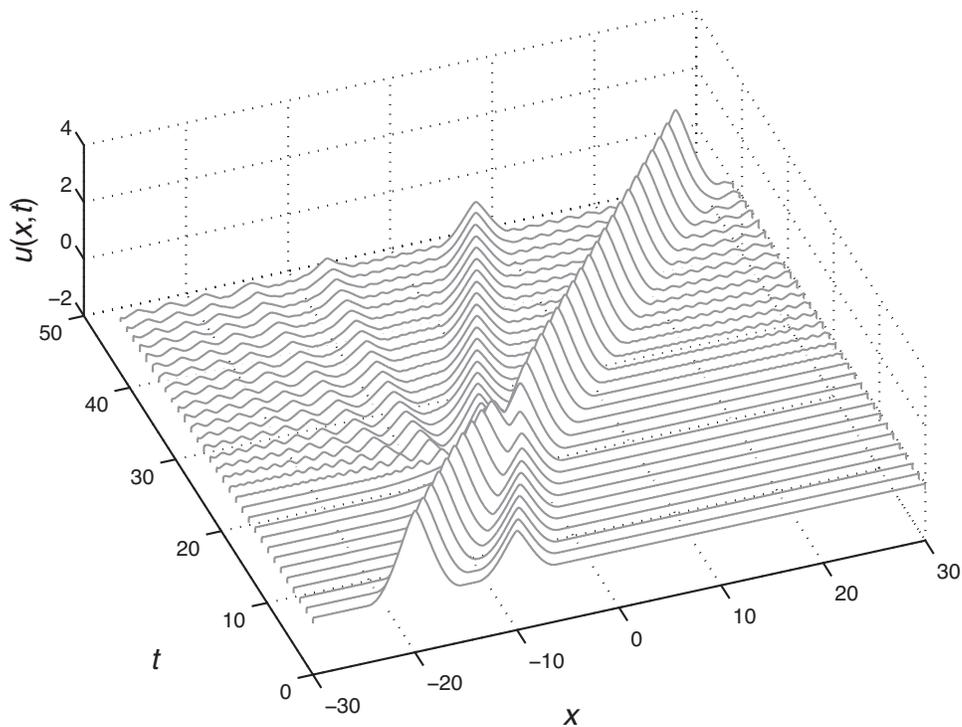}}
	\vspace*{8pt}
	\caption{Space-time plot of an overtaking collision of two peakompactons of the $K^\#(1,3)$ equation.}
	\label{fig:overtake}
\end{figure}

Our featured simulation is presented in Fig.~\ref{fig:overtake} in the form of a ``waterfall'' space-time plot. In this simulation, we have taken $L = 30$, $t_\mathrm{final} = 42$, $\Delta x = 0.05$ ($\Rightarrow N=1201$), $\Delta t$ is determined via the CFL condition \eqref{eq:CFL} with $C_\mathrm{CFL} = 0.1$, and $\nu = 3$. For the initial condition's parameters, we have used $\xi_1 = -10$, $c_1 = 0.5$, $\xi_2 = -20$ and $c_2 = 1$. Throughout the simulation, we monitored the evolution of the invariants $M$, $P$ and $H$ from Eq.~\eqref{eq:HMP} and found that $M$ is conserved within $0.03\%$. Meanwhile, $P$ and $H$ are conserved within $3.7\%$ and $4.7\%$, respectively. Furthermore, doubling the grid size did not change the qualitative features of the interaction, which we describe below. Hence, we have a degree of confidence in the quality and reproducibility of the numerical simulation shown in Fig.~\ref{fig:overtake}.

In the space-time plot presented in Fig.~\ref{fig:overtake}, we observe that the two peakompactons were initially placed so that the faster (therefore, taller) is on the left and the slower (therefore, shorter) is on the right at $t=0$. Since peakompactons propagate to the right, this setup results in an overtaking collision in which the taller collides with and advances past the shorter peakompacton. This is evident in the space-time plot by visually tracking the initial peaks. Clearly, the two peakompactons survive the collision and continue to propagate ``unharmed.'' However, the collision is not purely elastic as radiation modes (and possibly newly generated peakompactons) emerge from the moment of interaction, which occurs shortly after $t=20$. Aside from these radiation modes, the space-time plot in Fig.~\ref{fig:overtake} is quite typical of soliton collisions in the KdV equation (see, {\it e.g.}, Fig.~3.3 in Ref.~\refcite{Scott2003}). Thus, so far, we can conclude peakompactons appear to be stable objects that can collide and retain their identity. But, the presence of further propagation modes emerging from the collisions leads us to term the collision as \emph{nearly} elastic.

\begin{figure}[b]
	\centerline{\includegraphics[width=\textwidth]{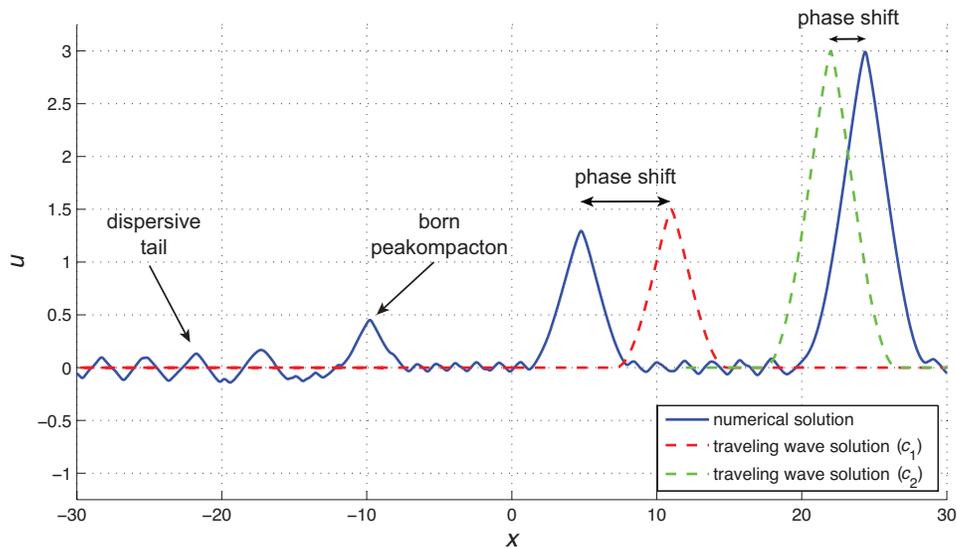}}
	\vspace*{8pt}
	\caption{Snapshot of the wavefield after an overtaking collision of two peakompactons. This figure shows the final $t$-trace from the spacetime plot in Fig.~\ref{fig:overtake} with  $u(x,t_\mathrm{final})$ as a solid curve and the locations of the original peakompactons, had each propagated by itself (undisturbed), as dashed curves. The phase shifts of the interacting peakompactons with respect to the undisturbed ones are highlighted as well as the birth of a third peakompacton and the generation of a self-similar dispersive tail (``Airy pulse'').}
	\label{fig:overtake2}
\end{figure}

To further elucidate the dynamics of the overtaking collision, in Fig.~\ref{fig:overtake2}, we show $u(x,t_\mathrm{final})$ as a function of $x$ (solid curve), while the would-be current location of each individual peakompacton, \emph{had it propagated by itself (undisturbed)}, is shown by a dashed curve. Clearly, the final locations of the interacting peakompactons differ from those of the undisturbed ones. This difference in location is termed a \emph{phase shift}. The taller one is ahead, while the shorter one is behind. Furthermore, while the taller peakompacton appears to have preserved its shape identically, the shorter peakompacton is markedly even shorter. We conjecture that this is due to the ``birth'' of a third, trailing peakompacton at the moment of interaction. Furthermore, the self-similar dispersive tail seen in Fig.~\ref{fig:overtake} is clearly depicted in Fig.~\ref{fig:overtake2}. The tail has apparently reached the left end of the periodic domain ($x=-L$), thus oscillations from it can be observed throughout the whole domain. The emergence of a self-similar dispersive pulse, which could be termed an Airy pulse, from the collision of two peakompactons is reminiscent of collisions in higher-order compacton equations\cite{Cooper2001} and of collisions of ``$\sech^2$'' pulses in the Korteweg--de Vries--Kuramoto--Sivashinsky--Velarde (KdV--KSV) equation,\cite{Christov1995} which is a higher-order KdV-like equation incorporating both viscous and hyperviscous terms as well as Marangoni effects. Finally, we do note that further numerical experiments suggest that the features of the collision just described, namely that (i) the peakompactons ``survive'' the collision, (ii) a third peakompacton is ``born'' and (iii) a dispersive tail is generated, are robust. 


\section{Conclusion}

In the present work, we have provided a comprehensive introduction to/tutorial on a class of nonlinearly dispersive KdV-like equations termed the $K^\#(n,m)$ equations. We reviewed the physical origin of these PDEs in the context of wave propagation in continua with microstructure. The Hamiltonian structure of the $K^\#(n,m)$ equations was examined, showing that they possess the usual three conservation laws of mass, momentum and energy. Significant effort was devoted to the construction of traveling wave solutions. Due to the nonlinearly dispersive nature of the PDEs, it was possible to construct compactly supported traveling waves, {\it i.e.}, waves of {finite} wavelength. These waveforms turned out to, additionally, exhibit certain derivative discontinuities at their crest, which led to the introduction of the term peakompacton. Several cases of the governing ODE for peakompactons were examined and analytical solutions presented together with the corresponding phase-plane analysis. Finally, we constructed a simple but highly reliable explicit three-level finite-difference scheme with filtered artificial viscosity to simulate the time evolution and collisions of peakompactons. We showed that peakompactons interact nearly elastically, preserving their general features in an overtaking collision. However, in what appears to be a hint of the non-integrability of the $K^\#(n,m)$ equations, peakompacton collisions can generate further peakompactons and dispersive pulses (``radiation''). Understanding the presence of such apparently self-similar radiation (see also Ref.~\refcite{Rus2007a}) emanating from the collision is a topic of future work.

Furthermore, even though we gathered some definitive information about collisions of peakompactons through numerical simulations, another avenue of future work in this direction is to use the {explicit} variational {\it ansatz} given in Eq.~\eqref{eq:anz2} to study collisions in the framework of the variational approximation, under which analytical results can be obtained. The calculation becomes significantly more difficult than the one we performed in Section~\ref{sec:va}. However, as shown for the case of the $\phi^4$ equation,\cite{Campbell1983} the $\phi^6$ equation,\cite{Gani2014} the sine-Gordon equation\cite{Christov2008,Ferguson1998} and the nonlinear Schr\"odinger equation,\cite{Baron2014} there are ways to extract important collision information from the coupled set of ODEs governing the parameters of the linear superposition of two independent instances of the variational {\it ansatz}.

Finally, an important topic that we have not mentioned at all in the present work is \emph{stability}\cite{Dey1998} of the constructed traveling wave solutions as functions of $n$, $m$, etc. Stability can be explored, to some degree, through numerical experimentation but much more can be said by analytical methods. We will not attempt to summarize the extensive literature on the topic in the limited space of the conclusion, beyond noting that this is an avenue of future work.


%
%
%

\section*{Acknowledgements}

This work was initiated while I.C.C.\ and T.K.\ were enjoying the hospitality of the Center for Nonlinear Studies at Los Alamos National Laboratory (LANL). I.C.C.\ was partially supported by the LANL/LDRD Program through a Feynman Distinguished Fellowship. LANL is operated by Los Alamos National Security, L.L.C.\ for the National Nuclear Security Administration of the U.S.\ Department of Energy under Contract No.\ DE-AC52-06NA25396.
I.C.C.\ would also like to thank F.\ Cooper, D.D.\ Holm, J.M.\ Hyman, P.M.\ Jordan and A.\ Oron for helpful discussions on the topic of the present work. 
Last but not least, we would like to thank Surajit Sen for organizing this special issue, for inviting us to contribute and for hosting I.C.C.\ and A.S.\ at the 2016 workshop on ``Nonlinear Dynamics of Many Body Systems'' in Buffalo, New York, where the idea for this manuscript was finalized.

%


\section*{References}

\begin{thebibliography}{999}








\bibitem{GenContA}
G.~A.~Maugin and A.~V.~Metrikine (eds.), {\it Mechanics of Generalized Continua: One Hundred Years After the Cosserats} (Springer Science+Business Media, New York, 2010).

\bibitem{GenContB}
H.~Altenbach, G.~A.~Maugin, V.~Erofeev (eds.), {\it Mechanics of Generalized Continua} (Springer-Verlag, Berlin/Heidelberg, 2011).

\bibitem{Rubin1995}
M.~B.~Rubin, P.~Rosenau and O.~Gottlieb, {\it J. Appl. Phys.} {\bf 77}, 4054 (1995).

\bibitem{LandauLifshitz}
L.~D.~Landau and E.~M.~Lifshitz, {\it Fluid Mechanics}, 2nd ed.\ (Pergamon Press, Oxford, UK, 1987).

\bibitem{CN60}
B.~D.~Coleman and W.~Noll, {\it Arch.\ Rational Mech.\ Anal.}\ {\bf 6}, 25 (1960).

\bibitem{DF74}
J.~E.~Dunn and R.~L.~Fosdick, {\it Arch.\ Rational Mech.\ Anal.}\ {\bf 56}, 191 (1974).

\bibitem{Chen}
S.~Chen, C.~Foias, D.~D.~Holm, E.~Olson, E.~S.~Titi and S.~Wynne, {\it Physica D} {\bf 133}, 49 (1999).

\bibitem{FHT01}
C.~Foias, D.~D.~Holm and E.~S.~Titi, {\it Physica D} {\bf 152--153}, 505 (2001).

\bibitem{GOP03}
J.~L.~Guermond, J.~T.~Oden and S.~Prudhomme, {\it Physica D} {\bf 177}, 23 (2003). 

\bibitem{Holm1998b}
D.~D.~Holm, J.~E.~Marsden and T.~S.~Ratiu, {\it Phys.\ Rev.\ Lett.}\ {\bf 80}, 4173 (1998).

\bibitem{Holm2002}
D.~D.~Holm, {\it Physica D} {\bf 170}, 253 (2002).

\bibitem{BF06}
H.~S.~Bhat and R.~C.~Fetecau, {\it Discret.\ Contin.\ Dyn.\ Syst.\ B} {\bf 6}, 979 (2006).

\bibitem{MKJC11}
R.~S.~Keiffer, R.~McNorton, P.~M.~Jordan and I.~C.~Christov, {\it Wave Motion} {\bf 48}, 782 (2011). 

\bibitem{Margolin2009}
L.~G.~Margolin, {\it Phil.\ Trans.\ R.\ Soc.\ A} {\bf 367}, 2861 (2009).

\bibitem{Margolin2014}
L.~G.~Margolin, {\it Mech.\ Res.\ Commun.}\ {\bf 57}, 10 (2014).

\bibitem{Jordan2015}
P.~M.~Jordan and R.~S.~Keiffer, {\it Phys.\ Lett.\ A} {\bf 379}, 124 (2015).

\bibitem{Margolin2016}
L.~G.~Margolin, in {\it Coarse Grained Simulation and Turbulent Mixing}, ed.\ F.~F.~Grinstein (Cambridge University Press, Cambridge, UK, 2016), pp.~48--86.

\bibitem{Destrade2006}
M.~Destrade and G.~Saccomandi, {\it Phys.\ Rev.\ E} {\bf 73}, 065604 (2006).

\bibitem{Jordan2012}
P.~M.~Jordan and G.~Saccomandi, {\it Proc.\ R.\ Soc.\ A} {\bf 468}, 3441 (2012).

\bibitem{Rubin2013}
M.~B.~Rubin, {\it Proc.\ R.\ Soc.\ A} {\bf 469}, 20120641 (2013).

\bibitem{Jeffrey1972}
A.~Jeffrey and T.~Kakutani, {\it SIAM Rev.}\ {\bf 14}, 582 (1972).

\bibitem{Crighton1979}
D.~G.~Crighton, {\it Annu.\ Rev.\ Fluid Mech.}\ {\bf 11}, 11 (1979).

\bibitem{Maugin2011}
G.~A. Maugin, {\it Mech.\ Res.\ Commun.}\ {\bf 38}, 341 (2011).

\bibitem{Laut2007}
W.~Lauterborn, T.~Kurz and I.~Akhatov, in {\it Springer Handbook of Acoustics}, ed.\ T.~D.~Rossing (Springer Science+Business Media, New York, 2007), pp.~257--297.

\bibitem{Jordan2014}
P.~M.~Jordan, R.~S.~Keiffer and G.~Saccomandi, {\it Wave Motion} {\bf 51}, 382 (2014).

\bibitem{Miura1968}
R.~M.~Miura, {\it J.\ Math.\ Phys.}\ {\bf 9}, 1202 (1968).

\bibitem{C15}
I.~C.~Christov, {\it Proc.\ Estonian Acad.\ Sci.}\ {\bf 64}, 212 (2015), arXiv:1501.01044 [math-ph].

\bibitem{Holm2009}
D.~D.~Holm, T.~Schmah and C.~Stoica {\it Geometric Mechanics and Symmetry: From Finite to Infinite Dimensions} (Oxford University Press, New York, 2009).

\bibitem{Morrison1998}
P.~J.~Morrison, {\it Rev.\ Mod.\ Phys.}\ {\bf 70}, 467 (1998).

\bibitem{Zakharov1997}
V.~E.~Zakharov and E.~A.~Kuznetsov, {\it Physics--Uspekhi} {\bf 40}, 1087 (1997).

\bibitem{Olver1980}
P.~J.~Olver, {\it Math.\ Proc.\ Camb.\ Phil.\ Soc.}\ {\bf 88}, 71 (1980).

\bibitem{Boussinesq1877}
J.~Boussinesq, {\it M\'emoires pr\'es\'entes par divers savants \`{a} l'Acad\'{e}mie des Sciences de l'Institut de France} {\bf XXIII}, 1 (1877).

\bibitem{Korteweg1895}
D.~J.~Korteweg and G.~de Vries, {\it Phil. Mag. (Ser. 5)} {\bf 39}, 422 (1895).

\bibitem{Zabusky1965}
N.~J. Zabusky and M.~D. Kruskal, {\it Phys.\ Rev.\ Lett.}\ {\bf 15}, 240 (1965).

\bibitem{Miura1976}
R.~M.~Miura. {\it SIAM Rev.}\ {\bf 18}, 412 (1976).

\bibitem{Drazin1989}
P.~G.~Drazin and R.~S.~Johnson, {\it Solitons: An Introduction} (Cambridge University Press, Cambridge, UK, 1989).

\bibitem{Scott2003}
A.~Scott, {\it Nonlinear Science: Emergence and Dynamics of Coherent Structures}, 2nd ed.\ (Oxford University Press, Oxford, UK, 2003).

\bibitem{Kruskal1975}
M.~Kruskal, in: {\it Dynamical Systems, Theory and Applications}, ed.\ J.~Moser (Springer-Verlag, Berlin/Heidelberg, 1975), pp.~310--354.

\bibitem{Gardner1971}
C.~S.~Gardner, {\it J.\ Math.\ Phys.}\ {\bf 12}, 1548 (1971).

\bibitem{Gelfand2000}
I.~M.~Gelfand and S.~V.~Fomin, {\it Calculus of Variations} (Dover Publications, Mineola, New York, 2000).

\bibitem{Kruskal1966}
M.~D.~Kruskal and N.~J.~Zabusky, {\it J.\ Math.\ Phys.}\ {\bf 7}, 1256 (1966).

\bibitem{Maugin2002}
G.~A.~Maugin and C.~I.~Christov, in {\it Selected Topics in Nonlinear Wave Mechanics}, ed.\ C.~I.~Christov and A.~Guran (Birkh\"auser, Boston, 2002), pp.~117--160.

\bibitem{LandauLifshitz2}
L.~D.~Landau and E.~M.~Lifshitz, {\it Mechanics}, 3rd ed.\ (Butterworth-Heinemann, Oxford, UK, 1976).

\bibitem{Cooper1993}
F.~Cooper, H.~Shepard and P.~Sodano, {\it Phys.\ Rev.\ E} {\bf 48}, 4027 (1993).

\bibitem{Rosenau1993}
P.~Rosenau and J.~M.~Hyman, {\it Phys.\ Rev.\ Lett.}\ {\bf 70}, 564 (1993).

\bibitem{Rosenau1997}
P.~Rosenau, {\it Phys.\ Lett.\ A} {\bf 230}, 305 (1997).

\bibitem{Rosenau1999}
P.~Rosenau and D.~Levy, {\it Phys.\ Lett.\ A} {\bf 252}, 297 (1999).

\bibitem{Rosenau2000}
P.~Rosenau, {\it Phys.\ Lett.\ A} {\bf 275}, 193 (2000).

\bibitem{Rosenau2005}
P.~Rosenau, {\it Not.\ AMS} {\bf 52}, 738 (2005).

\bibitem{Rosenau2006}
P.~Rosenau, {\it Phys.\ Lett.\ A} {\bf 356}, 44 (2006).

\bibitem{Rus2009}
F.~Rus and F.~R.~Villatoro, {\it Appl.\ Math.\ Comput.}\ {\bf 215}, 1838 (2009).

\bibitem{Rosenau2014}
P.~Rosenau and A.~Oron, {\it Commun.\ Nonlinear Sci.\ Numer.\ Simulat.}\ {\bf 19}, 1329 (2014).

\bibitem{Khare1993}
A.~Khare and F. Cooper, {\it Phys.\ Rev.\ E} {\bf 48}, 4843 (1993), arXiv:patt-sol/9307002.

\bibitem{Cooper2006}
F.~Cooper, A.~Khare and A.~Saxena, {\it Complexity} {\bf 11}, 30 (2006), arXiv:nlin/0508010 [nlin.PS].

\bibitem{Bender1998}
C.~M.~Bender and S.~Boettcher, {\it Phys.\ Rev.\ Lett.}\ {\bf 80}, 5243 (1998); 	arXiv:physics/9712001 [math-ph].

\bibitem{Bender2007}
C.~M.~Bender, D.~C.~Brody, J.-H.~Chen and E.~Furlan, {\it J.\ Phys.\ A: Math.\ Theor.}\ {\bf 40}, F153 (2007); arXiv:math-ph/0610003.

\bibitem{Bender2009}
C.~M.~Bender, F.~Cooper, A.~Khare, B.~Mihaila and A.~Saxena, {\it Pramana} {\bf 73}, 375 (2009), arXiv:0810.3460 [math-ph].

\bibitem{Assis2010}
P.~E.~G.~Assis and A.~Fring, {\it Pramana} {\bf 74}, 857 (2010), arXiv:0901.1267 [hep-th].

\bibitem{Davis1962}
H.~T.~Davis, {\it Introduction to Nonlinear Differential and Integral Equations} (Dover Publications, Mineola, New York, 1962).

\bibitem{DLMF}
A.~B.~Olde Daalhuis, 
in {\it NIST Digital Library of Mathematical Functions}, ed.~F.~W.~J.~Olver, D.~W.~Lozier, R.~F.~Boisvert and C.~W.~Clark, Chapter 15, \url{http://dlmf.nist.gov/}, release 1.0.13, 2016.

\bibitem{Coddington}
E.~A.~Coddington, {\it An Introduction to Ordinary Differential Equations}, (Dover Publications, Mineola, New York, 1989).

\bibitem{Saccomandi2004}
G.~Saccomandi, {\it Int.\ J.\ Non-Linear Mech.}\ {\bf 39}, 331 (2004).

\bibitem{Destrade2007}
M.~Destrade, G.~Gaeta and G.~Saccomandi, {\it Phys.\ Rev.\ E} {\bf 75}, 047601 (2007), arXiv:0711.4437 [physics.class-ph].

\bibitem{Destrade2009}
M.~Destrade, P.~M.~Jordan and G.~Saccomandi, {\it EPL} {\bf 87}, 48001 (2009); arXiv:1303.0953 [nlin.PS].

\bibitem{DLMF2}
R.~A.~Askey and A.~B.~Olde Daalhuis, 
in {\it NIST Digital Library of Mathematical Functions}, ed.~F.~W.~J.~Olver, D.~W.~Lozier, R.~F.~Boisvert and C.~W.~Clark, \S16.13, \url{http://dlmf.nist.gov/}, release 1.0.13, 2016.

\bibitem{Coleman1967}
B.~D.~Coleman and M.~E.~Gurtin, {\it Phys.\ Fluids} {\bf 10}, 1454 (1967).

\bibitem{Wei2013}
D. Wei and P.~M.~Jordan, {\it Int.\ J.\ Non-Linear Mech.}\ {\bf 48}, 72 (2013).

\bibitem{Camassa1993}
R.~Camassa and D.~D.~Holm, {\it Phys.\ Rev.\ Lett.}\ {\bf 71}, 1661 (1993).

\bibitem{Camassa1994}
R.~Camassa, D.~D.~Holm and J.~M.~Hyman, {\it Adv.\ Appl.\ Mech.}\ {\bf 31}, 1 (1994).

\bibitem{Boyd1997}
J.~P.~Boyd, {\it Appl.\ Math.\ Comput.}\ {\bf 81}, 173 (1997).

\bibitem{Camassa2003}
R.~Camassa, {\it Discr.\ Contin.\ Dyn.\ Syst.\ B} {\bf 3}, 115 (2003).

\bibitem{Holm1998}
D.~D.~Holm, J.~E.~Marsden and T.~S.~Ratiu, {\it Adv.\ Math.}\ {\bf 137}, 1 (1998).

\bibitem{Alber1995}
M.~S.~Alber, R.~Camassa, D.~D.~Holm and J.~E.~Marsden, {\it Proc.\ R.\ Soc.\ A} {\bf 450}, 677 (1995).

\bibitem{Olver1996}
P.~J.~Olver and P.~Rosenau, {\it Phys.\ Rev.\ E} {\bf 53}, 1900 (1996).

\bibitem{Hunter1994}
J.~K.~Hunter and Y.~Zheng, {\it Physica D} {\bf 79}, 361 (1994).

\bibitem{Hunter1991}
J.~K.~Hunter and R.~Saxton, {\it SIAM J.\ Appl.\ Math.}\ {\bf 51}, 1498 (1991).

\bibitem{Hereman1989}
W.~Hereman, P.~P.~Banerjee and M.~R.~Chatterjee, {\it J.\ Phys.\ A: Math.\ Gen.}\ {\bf 22}, 241 (1989).

\bibitem{Li1997}
Y.~A.~Li, P.~J.~Olver and P.~Rosenau, in {\it Nonlinear Theory of Generalized Functions}, eds.~
M.~Oberguggenberger, M.~Grosser, M.~Kunzinger and G.~Hormann (CRC Press, Boca Raton, FL, 1999), pp.~129--145.

\bibitem{LiOlver1997}
Y.~A.~Li and P.~J.~Olver, {\it Discr.\ Contin.\ Dyn.\ Syst.\ A} {\bf 3}, 419 (1997).

\bibitem{LiOlver1998}
Y.~A.~Li and P.~J.~Olver, {\it Discr.\ Contin.\ Dyn.\ Syst.\ A} {\bf 4}, 159 (1998).

\bibitem{Geyer2016}
A.~Geyer and V.~Ma\~{n}osa, {\it Nonlinear Anal.\ RWA} {\bf 31}, 57 (2016).

\bibitem{Sugiyama1979}
T.~Sugiyama, {\it Prog.\ Theor.\ Phys.}\ {\bf 61}, 1550 (1979).

\bibitem{Anderson1983}
D.~Anderson, {\it Phys.\ Rev.\ A} {\bf 27}, 3135 (1983).

\bibitem{Rice1983}
M.~J.~Rice, {\it Phys.\ Rev.\ B} {\bf 28}, 3587 (1983).

\bibitem{Campbell1983}
D.~K.~Campbell, J.~F.~Schonfeld and C.~A.~Wingate, {\it Physica D} {\bf 9}, 1 (1983).

\bibitem{Kaup2007}
D.~J.~Kaup and T.~K.~Vogel, {\it Phys.\ Lett.\ A} {\bf 362}, 289 (2007).

\bibitem{Chong2012}
C.~Chong, D.~E.~Pelinovsky and G.~Schneider, {\it Physica D} {\bf 241}, 115 (2012).

\bibitem{Christov2008}
I.~Christov and C.~I.~Christov, {\it Phys.\ Lett.\ A} {\bf 372}, 841 (2008); arXiv:nlin/0612005 [nlin.PS].

\bibitem{Cooper2001}
F.~Cooper, J.~M.~Hyman and A.~Khare, {\it Phys.\ Rev.\ E} {\bf 64}, 026608 (2001), arXiv:patt-sol/9704003.

\bibitem{IG}
R.~B.~Paris, 
in {\it NIST Digital Library of Mathematical Functions}, ed.~F.~W.~J.~Olver, D.~W.~Lozier, R.~F.~Boisvert and C.~W.~Clark, Chapter 8, \url{http://dlmf.nist.gov/}, release 1.0.13, 2016.

\bibitem{Rus2007b}
F.~Rus and F.~R.~Villatoro, {\it Math.\ Comput.\ Simulat.}\ {\bf 76}, 188 (2007).

\bibitem{Mihaila2010a}
B.~Mihaila, A.~Cardenas, F.~Cooper and A.~Saxena, {\it Phys.\ Rev.\ E} {\bf 81}, 056708 (2010). 

\bibitem{Mihaila2010b}
B.~Mihaila, A.~Cardenas, F.~Cooper and A.~Saxena, {\it Phys.\ Rev.\ E} {\bf 82}, 066702 (2010). 

\bibitem{Cardenas2011}
A.~Cardenas, B.~Mihaila, F.~Cooper and A.~Saxena, {\it Phys.\ Rev.\ E} {\bf 83}, 066705 (2011). 

\bibitem{Vlieg1971}
A.~C.~Vliegenthart, {\it J.\ Eng.\ Math.}\ {\bf 5}, 137 (1971).

\bibitem{Frutos1995}
J.~de Frutos, M.~A.~L\'opez-Marcos and J.~M.~Sanz-Serna, {\it J.\ Comput.\ Phys.}\ {\bf 120}, 248 (1995).

\bibitem{Ismail1998}
M.~S.~Ismail and T.~R.~Taha, {\it Math.\ Comput.\ Simulat.}\ {\bf 47}, 519 (1998).

\bibitem{Chertock2001}
A.~Chertock and D.~Levy, {\it J.\ Comput.\ Phys.}\ {\bf 171}, 708 (2001).

\bibitem{Levy2004}
D.~Levy, C.-W.~Shu and J.~Yan, {\it J.\ Comput.\ Phys.}\ {\bf 196}, 751 (2004).

\bibitem{Saucez2004}
P.~Saucez, A.~Vande Wouwer, W.~E.~Schiesser and P.~Zegeling, {\it J.\ Comput.\ Appl.\ Math.}\ {\bf 168}, 413 (2004).

\bibitem{Strikwerda2004}
J.~C.~Strikwerda, {\it Finite Difference Schemes and Partial Differential Equations}, 2nd ed.\ (Society for Industrial and Applied Mathematics, Philadelphia, 2004).

\bibitem{Lynch2005}
D.~R.~Lynch, {\it Numerical Partial Differential Equations for Environmental Scientists and Engineers} (Springer Science+Business Media, New York, 2005).

\bibitem{Rus2007a}
F.~Rus and F.~R.~Villatoro, {\it J.\ Comput.\ Phys.}\ {\bf 227}, 440 (2007); arXiv:0708.0486 [math-ph].

\bibitem{Rus2008}
F.~Rus and F.~R.~Villatoro, {\it App.\ Math.\ Comput.}\ {\bf 204}, 416 (2008).

\bibitem{Garralon2013}
J.~Garral\'{o}n, F.~Rus and F.~R.~Villatoro, {\it Appl.\ Math.\ Comput.}\ {\bf 220}, 185 (2013); arXiv:1209.1944 [math.NA].

\bibitem{Hyman1979}
J.~M.~Hyman, in {\it Advances in Computational Methods for PDEs--III}, ed.\ R.~Vichnevetsky and R.~S.~Stepleman (IMACS, 1979), pp.~313--321.

\bibitem{Christov1995}
C.~I.~Christov and M.~G.~Velarde, {\it Physica D} {\bf 86}, 323 (1995).

\bibitem{Gani2014}
V.~A.~Gani, A.~E.~Kudryavtsev and M.~A.~Lizunova, {\it Phys.\ Rev.\ D} {\bf 89}, 125009 (2014); arXiv:1402.5903 [hep-th].

\bibitem{Ferguson1998}
C.~D.~Ferguson and C.~R.~Willis, {\it Physica D} {\bf 119}, 283 (1998).

\bibitem{Baron2014}
H.~E.~Baron, G.~Luchini and W.~J.~Zakrzewski, {\it J.\ Phys.\ A: Math.\ Theor.}\ {\bf 47}, 265201 (2014); arXiv:1308.4072 [hep-th].

\bibitem{Dey1998}
B.~Dey and A.~Khare, {\it Phys.\ Rev.\ E} {\bf 58}, R2741 (1998).







  

%
%
%

\end{thebibliography}
\end{document}